\documentclass[apj]{emulateapj}
\usepackage{times}
\usepackage{amsfonts}
\usepackage{graphicx}

\newcommand{\etal}{{et al.~}}
\newcommand{\msunh}{\>h^{-1}\rm M_\odot}
\newcommand{\Msun}{\>{\rm M_\odot}}
\newcommand{\Lsunhh}{\,h^{-2}\rm L_\odot}

\newcommand{\mpch}{\>h^{-1}{\rm {Mpc}}}

\newcommand{\rmag}{\>^{0.1}{\rm M}_r-5\log h}

\usepackage{color}
\shorttitle{Member galaxy luminosities as halo mass proxies}

\shortauthors{Lu et al.}

\begin{document}

%%%%%%%%%%%%%%%%%%%%%%%%%%%%%%%%%%%%%%%%%%%%%%%%%%%%%%%%%%%%%%%%%%%%%%%%%%

\title{Using member galaxy luminosities as halo mass proxies of galaxy groups}

\author{Yi Lu\altaffilmark{1}, Xiaohu Yang\altaffilmark{1,2,3}, Shiyin
  Shen\altaffilmark{1}}

\altaffiltext{1}{Key Laboratory for Research in Galaxies and Cosmology,
  Shanghai Astronomical Observatory, Nandan Road 80, Shanghai 200030, China;
  E-mail:luyi@shao.ac.cn}

\altaffiltext{2}{Center for Astronomy and Astrophysics, Shanghai Jiao Tong
  University, Shanghai 200240, China; E-mail:xyang@sjtu.edu.cn}

\altaffiltext{3}{IFSA Collaborative Innovation Center, Shanghai Jiao Tong
University, Shanghai 200240, China}

%%%%%%%%%%%%%%%%%%%%%%%%%%%%%%%%%%%%%%%%%%%%%%%%%%%%%%%%%%%%%%%%%%%%%%%%%%

\begin{abstract}
  Reliable halo mass estimation for a given galaxy system plays an
  important role both in cosmology and galaxy formation studies.  Here
  we set out to find the way that can improve the halo mass estimation
  for those galaxy systems with limited brightest member galaxies been
  observed.  Using four mock galaxy samples constructed from
  semi-analytical formation models, the subhalo abundance matching
  method and the conditional luminosity functions, respectively, we
  find that the luminosity gap between the brightest and the
  subsequent brightest member galaxies in a halo (group) can be used
  to significantly reduce the scatter in the halo mass estimation
  based on the luminosity of the brightest galaxy alone. Tests show
  that these corrections can significantly reduce the scatter in the
  halo mass estimations by $\sim 50\%$ to $\sim 70\%$ in massive halos
  depending on which member galaxies are considered. Comparing to the
  traditional ranking method, we find that this method works better
  for groups with less than five members, or in observations with very
  bright magnitude cut.
\end{abstract}


\section[]{Introduction}

In the current scenario of galaxy formation, galaxies are thought to
be formed and reside in cold dark matter haloes. Studying the
galaxy-halo connection observationally provides one with important
information about the underlying processes in galaxy formation and
evolution. In recent years great progress has been made in
establishing the halo-galaxy link via the so called halo occupation
models (Jing, Mo \& B\"orner 1998; Mo, Mao \& White 1999; Peacock \&
Smith 2000; Zheng et al.2007) and the conditional luminosity functions
(Yang et al. 2003; van den Bosch et al. 2003; Cooray et al. 2006).
This halo-galaxy connection can also be made directly using galaxy
groups which are defined as sets of galaxies that reside in the same
dark matter halo, as is done in Yang et al. (2005c,d), Collister \&
Lahav (2005), van den Bosch et al. (2005), Robotham (2006), Zandivarez
et al. (2006), Weinmann et al. (2006a,b).

Numerous group catalogues have been constructed from galaxy redshift
surveys, including the 2-degree Field Galaxy Redshift Survey (2dFGRS)
(Eke et al. 2004; Yang et al.  2005a), the high-redshift DEEP2 survey
(e.g., Crook et al.2007); and the Sloan Digital Sky Survey (SDSS)
(e.g., Berlind et al. 2006; Yang et al.  2007; Tago et al.  2010;
Nurmi et al.  2013), the Galaxy and Mass Assembly (GAMA)
  observations (e.g. Robotham et al. 2011) using different methods,
ranging from the traditional friends-of-friends (FOF) algorithm, to
the hybrid matched filter method (Kim 2002) and the ``MaxBCG" method
(Koester et al.  2007). Relevant to the present paper is the
halo-based group finder developed by Yang et al. (2005a), which groups
galaxies according to their common halos expected from the current CDM
model.  This group finder is suited to study the relation between
galaxies and dark matter haloes over a wide dynamic range in halo
mass, from rich clusters of galaxies to poor groups of galaxies, as
tested with mock galaxy surveys, and has been applied to 2dFGRS (Yang
et al.  2005a), SDSS DR2 (Weinmann et al. 2006a), DR4 and DR7 (Yang et
al. 2007).

One of the key steps in the halo-based group finder is the estimation
of halo masses of candidate galaxy groups.  In general, the halo
masses of a group can be estimated based on the velocity dispersion of
their member galaxies.  However, except for a small fraction of very
rich groups with a large number of satellite galaxies (e.g., Carlberg
et al.  1996; 1997), this method is not very reliable for small groups
with only a few members. In general, stacking of satellite galaxies
for given central galaxies are used to obtain their halo masses (e.g.,
Erickson et al.  1987; Zaritsky et al.  1993; McKay et al.  2002;
Brainerd \& Specian 2003; Prada et al.  2003; van den Bosch et
al. 2004; Becker et al.  2007; Conroy et al.  2007; Norberg et al.
2008; More et al. 2009b, 2011; Li et al. 2012).

Group total luminosity (e.g.  Yang et al.  2005a; 2007) or richness
(e.g. Berlind et al. 2006; Becker et al. 2007; Andreon \& Hurn 2010;
Hao et al. 2010) can also be used as halo mass indicator.  As
demonstrated in Yang et al. (2007), the halo mass is tightly
correlated with the total luminosity of member galaxies.
Unfortunately not all member galaxies can be observed because galaxy
samples are usually magnitude limited.  In practice, one can only
measure a characteristic group luminosity, $L_G$, which is defined as
the total luminosity of member galaxies brighter than a given
luminosity threshold.  In the SDSS group catalogue of Yang et al.,
this luminosity threshold is chosen to be $\rmag=-19.5$, so that one
can reach to a depth of $z=0.09$. Here $\rmag$ is the absolute
$r$-magnitude, K- and E-corrected to $z=0.1$, the typical redshift of
galaxies in the SDSS redshift sample.  Assuming that $L_G$ increases
monotonically with halo mass $M_h$, one can then obtain the halo
masses of galaxy groups by abundance matching between groups and dark
matter halos,
\begin{equation}
\int_{L_G}^\infty n_G (L_G') dL_G'
=\int_{M_h}^\infty n_h (M_h') dM_h'\,,
\end{equation}
where $n_G(L_G)$ is the number density of groups at characteristic
luminosity $L_G$, and $n_h(M_h)$ is the halo mass function for the
cosmology adopted.  In Yang et al.  (2007), $M_h-L_G$ relation so
obtained is used iteratively in the group finder to link galaxies in
the same dark matter halo, and to assign halo masses to individual
galaxy groups.

As shown in Yang et al. (2005a; 2007) using mock galaxy catalogs, the
halo based group finder and the mass assignment scheme work well for a
moderately deep and large survey, such as the 2dFGRS and SDSS.
However, the method may not work as well for shallow surveys, such as
the 6dFGRS (Jones et al.  2009), and for groups selected from high
redshift surveys, such as DEEP2 (Willmer et al.  2006), COSMOS (Ilbert
et al.  2009) and GAMA (Driver et al. 2011) where only a few
(in most cases one or two) brightest member galaxies are observed in
each dark matter halo. In case the survey volume is very small or
difficult to calculate because of the bad survey geometry, the halo
mass estimation based on the ranking of $L_G$ may become unachievable.
Moreover, there are also some catalogues which only consists of a
certain type of galaxies, for example, the SDSS-III's Baryon
Oscillation Spectroscopic Survey (BOSS) (Dawson et al. 2013) which map
the spatial distribution of luminous red galaxies (LRGs). It is
unclear if the method can work well for those catalogues.  Thus, an
alternative estimate of the halo masses is required for such groups.

To estimate the halo masses for galaxy systems with only one or two
{\it brightest} member galaxies, one may make use of the central-host
halo relation obtained from conditional luminosity function (CLF; e.g.
Yang et al. 2003; van den Bosch et al. 2003), the subhalo abundance
matching (SHAM; e.g., Vale \& Ostriker 2006), or from predictions of
semi-analytical model (SAM; e.g., Kang et al.  2005).  As shown in
Yang et al.  (2008) and Cacciato et al.  (2009), the typical scatter
in $\log L_c$ for halos of a given mass is about 0.15. However for
massive halos $L_c\propto M_h^{\sim 0.25}$, and so the scatter in halo
mass for a given $L_c$ can be substantial.  Thus the central (or the
brightest) galaxy alone can not provide a reliable estimation of the
halo mass, especially for massive halos. Recently, suggestions have
been made to the magnitude gap between the brightest and second
brightest galaxies as another parameter to describe the halo mass in
addition to $L_c$ (e.g., Hearin et al. 2013a; More 2012; Shen et al.
2014).  In this paper, we investigate how halo masses are correlated
with group properties, such as the central galaxy\footnote{Throughout
  this paper, we refer to the brightest galaxy in a halo as the
  central galaxy.  } luminosity, the satellite member galaxy
luminosity or the luminosity gap, defined as $\log L_{\rm gap} = \log
L_c - \log L_s$, where $L_s$ is the luminosity of the i-th brightest
member galaxies.  For example, if a group only has two members, then
$L_s$ is the luminosity of the second brightest galaxy. If a group
contains five member galaxies, we choose to use the luminosity of the
fifth brightest member galaxy.  The goal of this paper is to improve
the halo mass estimation using both $L_c$ and the luminosity gap. The
investigations are carried out based on four mock samples generated
using SAM, SHAM, and two CLF models, respectively.

The outline of this paper is as follows.  \S \ref{sec_data} describes
the four mock samples used in this paper that are constructed using
the SAM, SHAM, and CLF, respectively.  In \S \ref{sec_mass} we present
the relationships between the halo mass and the total group (or
central galaxy) luminosity.  In \S \ref{sec_mass_gap} we show how the
luminosity gap affects the $L_c-M_h$ relation and can be used to
reduce the scatter within this relation.  In \S\ref{richness}, we
investigate the improvement of this reduction by involving more
fainter member galaxies in calculating luminosity gap.  In \S\ref{sec rank}, we
compare the performance of two halo mass estimation methods, which
are based on luminosity gap (refer to `GAP') and traditional ranking
method (`RANK') under different circumstances, respectively.  Finally,
we summarize our results in \S \ref{sec_conclusion}.

\section[]{samples of mock groups}
\label{sec_data}

In this paper, we make use of four sets of mock galaxy catalogs.  The
first is constructed by Guo et al.  (2011; hereafter G11) using a
semi-analytic model of galaxy formation applied to dark matter halo
merging trees obtained from the `Millennium Run' N-body simulation
(Springel et al.  2005).  The cosmological parameters adopted in the
simulation are $\Omega_m=0.25$, $\Omega_\Lambda=0.75$,
$h=0.73$ and a CDM
spectrum with an amplitude specified by $\sigma_8=0.9$. This set of
parameters is different from that obtained in recent observations
(e.g.  Planck Collaboration et al.  2013; Mandelbaum 2013). However,
for our test of accuracy of halo mass assignment, this is not a big
issue. The simulation was performed with GADGET2 (Springel 2005) using
$2160^3$ dark matter particles in a periodic cubic box with a side
length $L_{\rm box} = 500 \mpch$ (in comoving units).  The mass of a
particle is $8.6 \times 10^8 \msunh$.  In their modeling, G11 included
the growth and activity of the central black holes, as well as their
effects on suppressing the cooling and star formation in dark matter
halos.  Since we are interested in finding an accurate halo-mass proxy
from observed luminosities, here we only use the halo mass and
$r$-band luminosity of each galaxy given by G11.  We refer the reader
to G11 for the details of the semi-analytic model (see also in Guo et
al. 2013).  In this paper, we only make use of these galaxies with luminosity
$\log L \ga 8.0$.  This set of mock galaxy catalog is referred to as
`SAM'.

The second mock galaxy catalog is constructed by Hearin et al. (2013b).
They used the Bolshoi N-body simulation (Klypin et al.  2011) which
models the cosmological growth of structure in a cubic volume with
side length $L_{\rm box} =250 \mpch$ within a standard $\Lambda$CDM
cosmology with parameters: $\Omega_m=0.27$, $\Omega_\Lambda=0.73$, $
h=0.7$, $n=0.95$ and $\sigma_8=0.82$.  After utilizing the ROCKSTAR
(Behroozi et al.  2012) halo finder to identify halos and subhalos,
subhalo abundance matching (SHAM) technique was adopt to associate
galaxies with dark matter halos.  This technique, called as age
distribution matching, is a two-phase algorithm for assigning both
luminosity and color to the galaxies located at the center of halos
(see Hearin et al.  2013a; 2013b for details).  Here, we make use of
their galaxy catalog in the cubic box that contains 244784 galaxies at
redshift $z = 0.1$ with $r$-band absolute magnitude $\rmag < -19.0$.

The third mock galaxy sample used in this paper is constructed by
populating dark matter halos obtained from numerical simulations with
galaxies of different luminosities according to the conditional
luminosity function (CLF; Yang et al. 2003). The algorithm of
populating galaxies is similar to that outlined in Yang et al.  (2004)
but here updated to the CLF in $r$-band.  The CLF, $\Phi(L|M_h)$, is
defined to be the average number of galaxies of luminosities $L \pm
dL/2$ that reside in a halo of mass $M_h$ (see Yang et al.  2003).
Here we give a brief description of the algorithm of assigning
luminosity to each member galaxy. First, we write the total CLF as the
sum of the CLFs of central and satellite galaxies (Yang et
  al. 2003; van den Bosch et al. 2003; Vale \& Ostriker 2004, 2006;
  Cooray 2006; van den Bosch et al. 2007; Yang et al. 2008):
\begin{equation}\label{eq:CLF_fit}
\Phi(L|M_h) = \Phi_{\rm cen}(L|M_h) + \Phi_{\rm sat}(L|M_h)\,.
\end{equation}
We assume the contribution from the central galaxies to be a lognormal
distribution:
\begin{eqnarray}\label{eq:phi_c}
\Phi_{\rm cen}(L|M_h)  ~d \log L
 ~~~~~~~~~~~~~~~~~~~~~~~~~~~~~~~~~~~~~~~~~~~~~~~~~~~~~~~~~~~~ && \\
=  {1 \over {\sqrt{2\pi}\sigma_c}} {\rm exp}
\left[- { {(\log L  -\log L_c )^2 } \over 2\sigma_c^2} \right] ~d \log L
\nonumber \,, &&
\end{eqnarray}
where $\sigma_c$ is a free parameter, which expresses the scatter in
$\log L$ of central galaxies at fixed halo mass, and $\log L_c$ is the
expectation value for the (10-based) logarithm of the luminosity of
the central galaxy.  For the contribution from the satellite galaxies
we adopt a modified Schechter function which decreases faster at the
bright end than the Schechter function:
\begin{eqnarray}\label{eq:phi_s}
\Phi_{\rm sat}(L|M) ~d \log L
~~~~~~~~~~~~~~~~~~~~~~~~~~~~~~~~~~~~~~~~~~~~~~~~~~~~~~~~~~~~ && \\
= \phi^*_s
\left ( {L\over L^*_s}\right )^{(\alpha_s+1)} {\rm exp} \left[- \left
  ({L\over L^*_s}\right )^2 \right] \ln(10) ~d \log L \nonumber \,. &&
\end{eqnarray}
The parameters $L_c$, $\sigma_c$, $\phi^*_s$, $\alpha_s$ and $L^*_s$
are all functions of the halo mass $M_h$.

Following Yang et al. (2008) and Cacciato et al.  (2009), we assume
that $\sigma_c$ is a constant independent of halo mass, and that the
$L_c-M_h$ relation has the following functional form,
\begin{equation}\label{eq:Lc_fit}
L_c (M_h) = L_0 \frac { (M_h/M_1)^{\gamma_1} }{(1+M_h/M_1)^{\gamma_1-\gamma_2} } \,.
\end{equation}
This model contains four free parameters: a normalized luminosity,
$L_0$, a characteristic halo mass, $M_1$, and two slopes, $\gamma_1$
and $\gamma_2$.  For satellite galaxies we use
\begin{equation}
\log L^*_s(M_h)  = \log L_c(M_h) - 0.25\,,
\end{equation}
\begin{equation}\label{alpha}
\alpha_s(M_h) = \alpha_s
\end{equation}
(i.e., the faint-end slope of $\Phi_{\rm sat} (L|M_h)$ is independent
of halo mass), and
\begin{equation}\label{phi}
\log[\phi^*_s(M_h)] = b_0 + b_1 (\log M_{12}) + b_2 (\log M_{12})^2\,,
\end{equation}
with $M_{12}=M_h/(10^{12} h^{-1}\Msun)$. This set of CLF
parameterization thus has a total of nine free parameters,
characterized by the vector
\begin{equation}\label{lambdaCLF}
{\lambda}^{\rm CLF} \equiv (\log M_1, \log L_{0}, \gamma_1,
\gamma_2, \sigma_c, \alpha_s, b_0, b_1, b_2) \, .
\end{equation}
The above functional forms do not have ample physical motivations but
were found to adequately describe the observational results obtained
by Yang \etal (2008) from the SDSS galaxy group catalog. They were
adopted in Cacciato et al. (2009) to make the model prediction for the
galaxy-galaxy lensing signals, and very recently in van den Bosch et
al. (2013), More et al.  (2013) and Cacciato et al.  (2013) to
constrain both the CLF and cosmological parameters with the SDSS
clustering and weak lensing measurements.  Here we adopt the set of
the best fit CLF parameters that are obtained for the Millennium
simulation cosmology, with $(\log M_1 = 10.954, \log L_{0} = 9.896,
\gamma_1 = 5.192, \gamma_2 = 0.2415, \sigma_c = 0.1501,
\alpha_s=-0.6828, b_0 = -0.1611 , b_1 = 0.7945, b_2 = -0.1573)$.

Galaxies with luminosities $\log (L/h^{-2}{\rm L}_\odot) \ga 8.0$ are
populated in the dark matter halos extracted from the Millennium
Simulation.  In practice, each halo is assigned a central galaxy with
a median luminosity specified by Eq.\,(\ref{eq:Lc_fit}) and a
log-normal dispersion $\sigma_c=0.1501$.  The satellite galaxies are
populated via following steps: (1) obtain the mean number of satellite
galaxies according to the integration of Eq.(\ref{eq:phi_s}) with
luminosities $\log L \ga 8.0$; (2) use a Possion distribution with the
mean obtained in step (1) to obtain the number of satellite galaxies;
(3) assign a luminosity to each of these satellite galaxy according to
Eq.(\ref{eq:phi_s}). Note that satellite galaxies are allowed to be
brighter than the central galaxy. However, we note that in our
subsequent analysis, following the observational definition, the
central galaxy is defined as the brightest galaxy in a halo. We refer
to this set of mock galaxy catalog as `CLF1'.

Finally, in order to test if our analysis is cosmological dependent or
not, we construct another set of mock galaxy catalog which we call as
`CLF2'.  The cosmological parameters adopted for this set of mock
galaxy catalog are those given by WMAP3:
$\Omega_m=0.238$, $\Omega_\Lambda=0.762$, $
h=0.73$, $n=0.95$ and $\sigma_8=0.75$.  We adopt the set of the best
fit CLF parameters that are consistent with WMAP3 cosmology, provided
in Cacciato et al.  (2009), with $(\log M_1 = 11.07, \log L_{0} =
9.935, \gamma_1 = 3.273, \gamma_2 = 0.255, \sigma_c = 0.143)$ (see
Cacciato et al.  2009 for more details).  Here again, we use the halo
catalog from the `Millennium Run' N-body simulation to construct the
mock catalog.  Since the cosmology adopted in the Millennium
simulation is different from the WMAP3 cosmology, we scale the halo
masses in the Millennium simulation to the WMAP3 cosmology using
abundance matching according to the mass functions of WMAP3 cosmology
and the Millennium simulation.  Such abundance matching ignores the
difference in the spatial correlations between halos in different
cosmologies, but should be valid for our analysis which is based only
on galaxy occupation in individual halos.

Among the above four mock catalogs, although the SAM has more
`physics' in it, but doesn't really fit the data (e.g., see Weinmann
et al. 2006; Liu et al. 2010; Nurmi et al. 2013), whereas the CLF1,
CLF2 and SHAM mocks are not based on physical models of galaxy
formation but do fit the observational data.  Here in our subsequent
probes, we make use of CLF1 as our fiducial sample, and take others
mainly for comparisons.

\section[]{Halo masses from luminosities of member galaxies}
\label{sec_mass}

\begin{figure*}
\center
\vspace{0.5cm}
\includegraphics[height=10.0cm,width=11.0cm,angle=0]{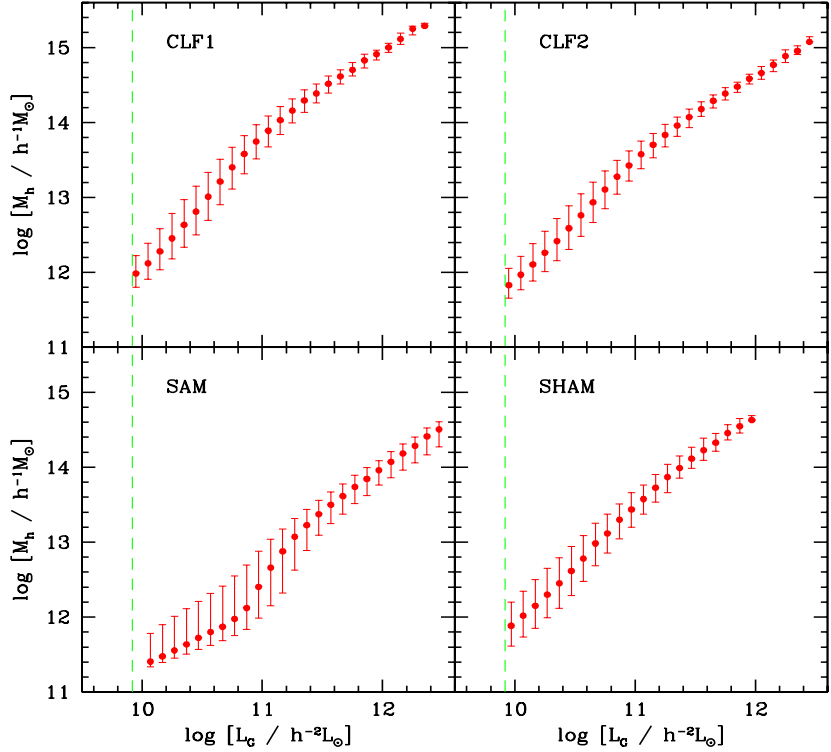}
\caption{The halo mass $M_h$ v.s.  characteristic group luminosity
  $L_G$ in the CLF1 (top left panel), CLF2 (top right panel), SAM
  (bottom left panel) and SHAM (bottom right panel) mock
  catalogs. Here results are shown for the median (symbols) and 68\%
  confidence level (error bars) of $\log M_h$ within each $\log L_G$
  bin.  The vertical green dashed line in each panel shows the
  luminosity limit at $\rmag=-19.5$. }
\label{fig:M-Lg}
\end{figure*}

In order to carry out cosmological and structure formation
investigations using observations of galaxy systems, one of the key
steps is to obtain reliable estimations of the halo masses of the
systems selected.  The kinematics of member galaxies is one of the
most commonly used methods to estimate masses of optically selected
groups/clusters (e.g., van den Bosch et al. 2004; Kargert et
  al. 2004; More et al. 2011). However, this method may be
significantly affected by anisotropies of the velocity dispersion
(e.g.  Biviano et al, 2006; White et al.  2010).  Furthermore, for
groups containing only a few members, or when only a few group members
can be observed, the estimation of velocity dispersion becomes noisy,
making the dynamical mass estimation unreliable.

Recently, another halo mass indicator has been used to estimate halo
masses for galaxy systems over a large dynamical range.  This method
uses the richness or the characteristic group luminosity as the proxy
of halo mass, and so can readily be applied once galaxy systems are
selected from a redshift survey.  Richness is one of the easiest
quantities of a galaxy system to observe, but there are different ways
to define the richness.  For instance, one can define it as the number
of member galaxies within the red sequence or above a certain
luminosity threshold.  Gladders \& Yee (2000, 2005) used the red
sequence richness to measure the mass of the halos from photometric
data.  The characteristic group luminosity (or stellar mass), which is
defined as the total group luminosity (stellar mass) of member
galaxies above a certain luminosity threshold, is presumably a better
mass indicator than the richness. This is especially the case for very
poor galaxy systems dominated by a single luminous galaxy, where the
galaxy luminosity is known to change as a function of halo mass over a
large range.

\begin{figure*}
\center
\vspace{0.5cm}
\includegraphics[height=13.0cm,width=11.0cm,angle=0]{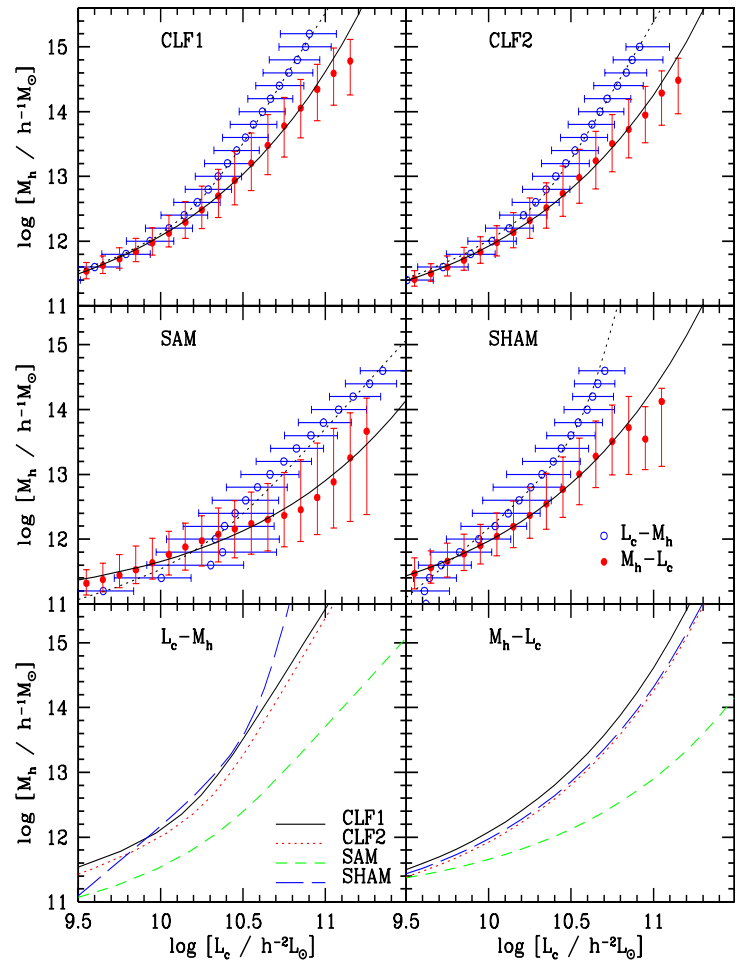}
\caption{The median halo mass, $M_h$, as function of central galaxy
  luminosity, $L_c$.  The red solid circles show the median values of
  $M_h$ in given central galaxy luminosity bins.  Error bars indicate
  the 68\% confidence level around the median values.  Here results
  are shown for mock samples CLF1 (top left), CLF2 (top right), SAM
  (middle left) and SHAM (middle right), respectively.  The solid line
  in each panel is the best fit $M_h-L_c$ relation.  As a comparison,
  we also show using blue open circles the median $\log L_c$ in different
  $\log M_h$ bins with horizontal error bars indicating the 68\%
  confidence levels around the median.  The dotted line in each panel
  is the best fit $L_c-M_h$ relation. The bottom two panels show all the
  fitting lines of $L_c-M_h$ relation (bottom left) and $M_h-L_c$
  relation (bottom right) for four mock samples respectively. }
\label{fig:Lc-Mh}
\end{figure*}

\subsection{Using the characteristic group luminosity}
\label{sec:LG-Mh}

As demonstrated in Yang et al.  (2005a; 2007) using mock redshift
surveys, the total luminosity of all member galaxies in a group is
tightly correlated with halo mass.  This method needs the
determination of the total group luminosity.  However, a limitation of
this method is the need of corrections in real observations where only
bright galaxies are observed.  Although one may make corrections using
a luminosity function (e.g. that for the total galaxy population),
detailed CLF modeling shows that the correction depends significantly
on halo mass (Yang et al. 2003; van den Bosch et al. 2007; Yang
  et al. 2008).  To avoid this uncertainty and as a compromise with
observational magnitude limit, one may use a characteristic group
luminosity, defined as the total luminosity of member galaxies above a
luminosity threshold.  For instance, Yang et al.  (2005a; 2007) used a
characteristic group luminosity, $L_G$, which is defined as the total
luminosity of all group members brighter than $^{0.1}{\rm
  M}_{b_J}-5\log h=-18$ for the 2dFGRS group catalogue, or of all
group members brighter than $\rmag =-19.5$ for the SDSS DR7 group
catalogue, as the halo mass proxy.  Tests using mock galaxy and group
catalogues show that the characteristic group luminosity is tightly
correlated with the mass of the dark matter halo hosting the group
(Yang et al. 2005a).  Once the characteristic group luminosities of
all the groups are obtained, one can assign groups with halo masses
according to abundance matching (e.g.  Mo, Mao \& White 1999) between
the group luminosity function and the halo mass function predicted
with a given cosmology (see Yang et al. 2007 for details).

Fig.  \ref{fig:M-Lg} shows the relation between the characteristic
group luminosity $L_G$ (the total luminosity of all group members
brighter than $\rmag =-19.5$) and the halo mass $M_h$ obtained
directly from the CLF1, CLF2, SAM and SHAM mock galaxy catalogs in
different panels as indicated, respectively. The symbols in each panel
are the median values while the error-bars are the 1-$\sigma$
variation (68\% confidence level).  There is a clear tight correlation
between the characteristic luminosity and the halo mass.  The typical
1-$\sigma$ variations in $\log M_h$ given by the SAM sample are $\sim
0.2$ for massive halos and $\sim 0.4$ for intermediate mass ($M_h\sim
10^{12.5}\msunh$) halos.  The corresponding variations given by the
CLF1, CLF2 and SHAM samples are $\sim 0.1$ for massive halos and $\sim
0.3$ in the intermediate mass range. Note that, the very
  asymmetric error bars at low luminosity end for SAM sample are
  caused by a combination of the color modeling of central galaxies in
  halos around $10^{11.8}\msunh$ (see Fig. 2 for such a feature for
  central galaxies), the luminosity cut at $\rmag=-19.5$ and a smaller
  bin size used here.

\subsection{Using the brightest (central) galaxy}

\begin{figure*}
\center
\vspace{0.5cm}
\includegraphics[height=10.0cm,width=11.0cm,angle=0]{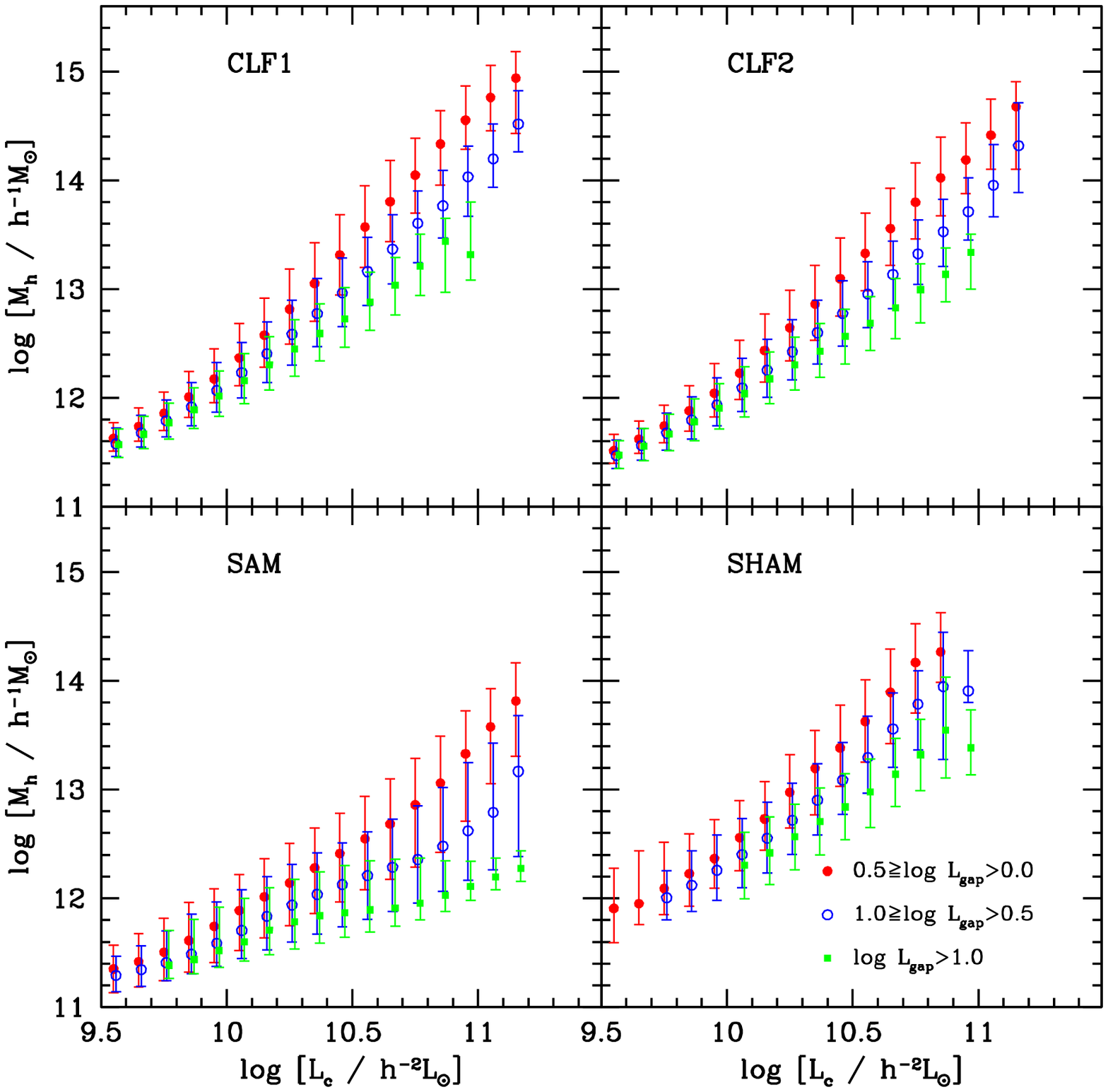}
\caption{Similar to the results shown in Fig.  \ref{fig:Lc-Mh}, but
  here for groups within three different luminosity gap ranges: $0.5
  \ge \log L_{\rm gap} > 0.0$ (solid circles), $1.0 \ge \log L_{\rm
    gap} > 0.5$ (open circles) and $\log L_{\rm gap} > 1.0$ (solid
  squares). Here again, symbols and error bars represent the median
  and 68\% confidence levels of $\log M_h$ in different $\log L_c$
  bins. }
\label{fig:Mh-gap}
\end{figure*}

In the local Universe, the characteristic group luminosities can be
determined for groups down to a halo mass of $\sim 10^{12}\msunh$ with
surveys like the 2dFGRS and the SDSS. However, if one goes to higher
redshift or uses much shallower observations (e.g. 6dFGRS), where only
a few brightest members a group can be observed, corrections for the
missing members have to be made to obtain the characteristic group
luminosity, making the method unreliable. On the other hand,
investigations with the CLF model and observed group and cluster
catalogues have shown that the luminosity of the central galaxy $L_c$
of a group is correlated with its halo mass $M_h$ (e.g., Yang et al.
2003; van den Bosch et al.  2003; Yang et al.  2007).  As an
illustration, Fig.\ref{fig:Lc-Mh} shows the model predictions of the
$L_c- M_h$ relation from the four mock samples described in Section
\ref{sec_data} as marked in each panel, respectively.  In each panel,
the blue open circles are the median values of the central galaxy
luminosity in different halo mass bins, and the horizontal error bars
show the 68\% confidence level around the median luminosities.  As one
can see, the median luminosity of central galaxy increases
monotonically with halo mass, and that the amounts of scatter are
comparable among four mock catalogs.  For reference, we fit the
$L_c-M_h$ relations using the functional form given by
Eq.\,(\ref{eq:Lc_fit}).  The fitting parameters are listed in Table
\ref{tab:Lc-Mh} and the best fitting results are shown in Fig.
\ref{fig:Lc-Mh} with the dotted lines.  For the CLF1 and CLF2 samples,
the parameters are adopted the same as those listed in Section
\ref{sec_data}.  These fitting formulae describe the median $L_c-M_h$
relations remarkably well.  For the most massive halos, the SAM
predicts central galaxies are significantly brighter than the CLF
model. However, as compared among SAMs and SDSS observations in G11,
the stellar mass functions at massive end of SAMs do not show very
significant deviation, the above difference may due to the color
modeling of these galaxies. Their model central galaxies,
  especially in $\sim 10^{11.8}\msunh$ halos, tend to be too blue and
  bright. This is the case of the wiggle in their $ M_h-L_c$ relation
  and the resulting much smaller median $L_c-M_h$ relation.

\begin{deluxetable}{lcccc}
 \tabletypesize{\scriptsize}
  \tablecaption{Parameters for the $L_c-M_h$ relation}
  \tablewidth{0pt}
  \tablehead{Sample & $\log L_0$ & $\log M_1$ & $\gamma_1$ & $\gamma_2$ } \\
 \startdata
CLF1     & 9.896 & 10.954 & 5.192 & 0.2415 \\
CLF2     & 9.935 & 11.07  & 3.273 & 0.255 \\
SAM      & 10.02$\pm$0.25 & 10.95$\pm$0.48 & 2.59$\pm$0.40 & 0.36$\pm$0.03\\
SHAM     & 10.25$\pm$0.16 & 12.94$\pm$0.56 & 0.36$\pm$0.03 & 0.22$\pm$0.03
\enddata
\label{tab:Lc-Mh}
\end{deluxetable}

Although the $L_c- M_h$ relations presented above are useful in
predicting the luminosities of central galaxies in halos of given
masses, these relations are not appropriate for estimating halo mass
from $L_c$.  Because the luminosity function is steep at the bright
end, and because the scatter in the $L_c(M_h)$ relations is quite
large, $M_h$ will be over-estimated from the $L_c(M_h)$ relation due
to a Malmquist-like bias. In order to have an unbiased result, we need
to obtain the $M_h(L_c)$ relation directly instead of using the
inverse of the $L_c(M_h)$ relation.  In Fig.  \ref{fig:Lc-Mh} we plot
the median (red solid circles) and the 68\% confidence levels (error bars)
of halo masses, $\log M_h$, as a function of central galaxy luminosity
$\log L_c$, obtained directly from the CLF1, CLF2, SAM and SHAM
samples.  Here again the median halo mass increases monotonically with
the central galaxy luminosity.  For low-mass halos, the median of the
$M_h (L_c)$ relation is similar to that of the $L_c(M_h)$ relation,
but for halos with mass $M_h>10^{12}\msunh$ these two relations are
quite different, exactly because of the Malmquist-like bias.  The
scatter in $\log M_h$ for a given luminosity increases with $\log
L_c$, and at the bright end is much larger than that in the $L_c(M_h)$
relation, particularly in the SAM prediction, clearly due to the
shallow slope in the $L_c(M_h)$ relation (for example, $L_c \propto
M_h^{\sim 0.25}$ at the massive end for the CLF sample) (see More et
al.  2009a; 2009b for more detailed discussions).  On the other hand,
as pointed out in More et al.  (2009b), the huge scatter in $M_h$ at
large $L_c$ seen in SAM is inconsistent with the constraints obtained
using satellite kinematics.  We fit the median $M_h(L_c)$ relations
with the following functional form:
\begin{equation}\label{eq:Mh-Lc}
\log M_h = \exp(\log L_c-\log M_a)+ \log M_b \,.
\end{equation}
The best fitting parameters of $[\log M_a, \log M_b]$ are: $[9.61 \pm
0.01 , 10.60 \pm 0.01]$ for the CLF1 sample, $[9.69 \pm 0.02 , 10.56
\pm 0.03]$ for the CLF2 sample, $[10.32 \pm 0.03 , 10.93 \pm 0.04]$
for SAM sample and $[9.68 \pm 0.02 , 10.64 \pm 0.03]$ for SHAM sample,
respectively. The resulting best fit median relations are shown in
Fig. 2 as the solid lines. Note that the above fitting formula
  are obtained within the halo mass range shown in the plot,
  extrapolation to much larger halo mass is not valid.  For clarity,
  we put all the fitting lines of four mock samples in the bottom two
  panels for $L_c- M_h$ relations on the right and $M_h- L_c$
  relations on the left. Obviously, for SAM, both relations are
  inconsistent with the other three mock samples. As we have mentioned
  before, it's because the galaxy luminosity in SAM model is
  overestimated compare to the other three mock samples. 

The above results show that the scatter in the $M_h(L_c)$ is quite
large at the massive end. It is therefore not a good choice to use
$L_c$ alone as a halo mass proxy. On the other hand, we see in Fig. 1
that the scatter shown for a given characteristic group luminosity is
much smaller, which indicates that using additional member galaxies
can give better constrain of the halo masses.  In what follows we
investigate how to get a better halo mass proxy by using the
luminosities of other brightest member galaxies.

\subsection[]{Using the brightest two galaxies}
\label{sec_mass_gap}

\begin{figure*}
\center
\vspace{0.5cm}
\includegraphics[height=10.0cm,width=11.0cm,angle=0]{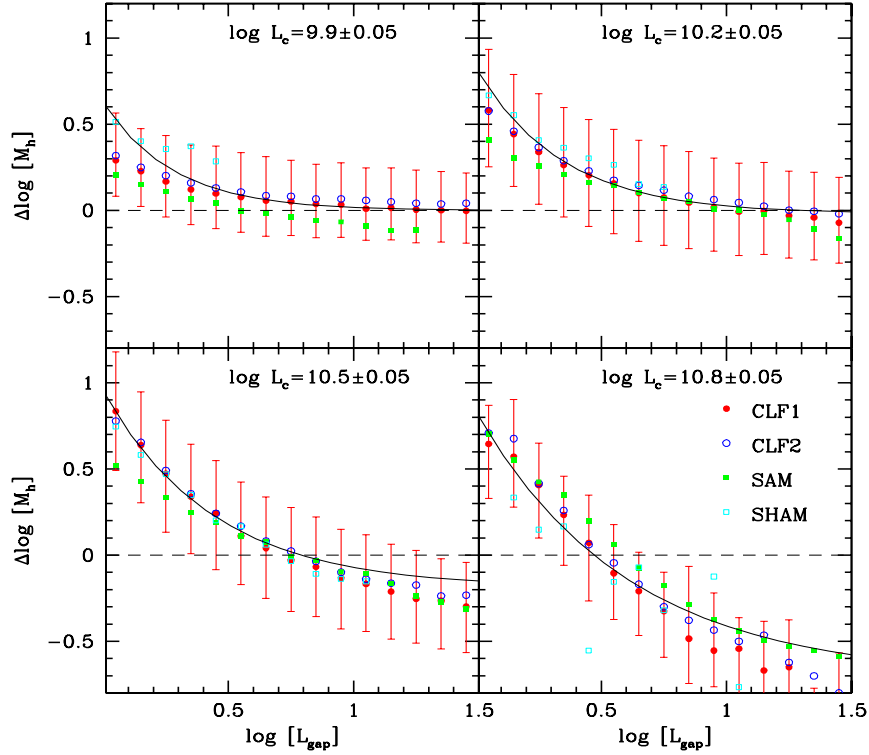}
\caption{The halo mass difference, $\Delta \log M_h$, between the best
  fit $M_h - L_c$ relation and data as a function of luminosity gap
  $\log L_{\rm gap}$.  The results shown in different panels are for
  central galaxies within different luminosity bins: $\log L_c =
  9.9\pm 0.05$ (upper-left), $10.2\pm 0.05$ (upper-right), $10.5\pm
  0.05$ (lower-left) and $10.8\pm 0.05$ (lower-right), respectively.
  In each panel, red solid circles are results obtained from the CLF1
  sample, where the symbols and error bars represent the median and
  68\% confidence levels of $\Delta \log M_h$.  Black solid line is
  the best fitting results to the data from CLF1 sample. The results
  obtained from CLF2, SAM and SHAM samples are also shown in each
  panel for comparison using different symbols as indicated. }
\label{fig:DeltaM-gap}
\end{figure*}

To start with, we introduce a `luminosity gap', defined as the
luminosity ratio between the central and a satellite galaxy in the same dark
matter halo, $L_{\rm gap}= L_c/L_s$.  In this section, we focus on the
brightest satellite galaxy, so that $L_s=L_2$, where $L_2$ is the
luminosity of the brightest satellite or the second brightest among
all the member galaxies.  Recent studies have shown that this
luminosity gap may contain important information about halo mass
(e.g., Hearin et al.  2013; More 2012).  As found in Shen et al.
(2014), $L_2$ is also correlated with group richness which, in turn,
is correlated with halo mass.  Thus a combination of the luminosities
of the brightest and second brightest galaxies in a halo is likely to provide
more information regarding the halo mass than the brightest galaxy
alone.

The luminosity gap $L_{\rm gap}$ can be measured straightforwardly for
groups/halos that have at least two members.  Within our four samples,
the distribution of the luminosity gap, $\log L_{\rm gap}$, spreads
roughly in the range 0.0-3.0. We split each of our mock group
catalogues into three subsamples according to the value of $\log
L_{\rm gap}$: small gap groups with $0.5 \ge \log L_{\rm gap} > 0.0$;
median gap groups with $1.0 \ge \log L_{\rm gap} > 0.5$; and large gap
groups with $\log L_{\rm gap} > 1.0$.  Fig. \ref{fig:Mh-gap} shows the
same $M_h( L_c)$ relations as shown in Fig.  \ref{fig:Lc-Mh}, but
separately for each of the gap subsamples, as indicated by different
symbols.  Clearly, groups with different amounts of luminosity gap
have distinctive $M_h(L_c)$ relations, especially at the massive end.
For the same central galaxy luminosity, groups with a smaller
luminosity gap (or larger $L_2$) tends to be more massive, especially
for groups whose central galaxy luminosities are larger than
$10^{10.0} \Lsunhh$.  For fainter central galaxies, the $M_h(L_c)$
relations are similar for the three luminosity gaps, due to the small
scatter in the halo mass - central galaxy luminosity relation.
We may notice the much larger reduction of error bars at bright
  central luminosity end for large gap groups with $\log L_{\rm gap} >
  1.0$ in SAM mock sample. This is also due to the fact that in SAM
  the luminosities of central galaxies in a large number of low mass
  groups are overestimated compared to the other three mock
  samples. As we mentioned before, this overestimation of luminosity
  may come from the incorrect color modeling of these galaxies. 

The strength of the luminosity-gap dependence increases {\it
  monotonically} with the central galaxy luminosity. All these suggest
that the luminosity gap can be used to reduce the scatter in the
$M_h-L_c$ relations, particularly at the massive end.

\begin{figure*}
\center \vspace{0.5cm}
\includegraphics[height=10.0cm,width=11.0cm,angle=0]{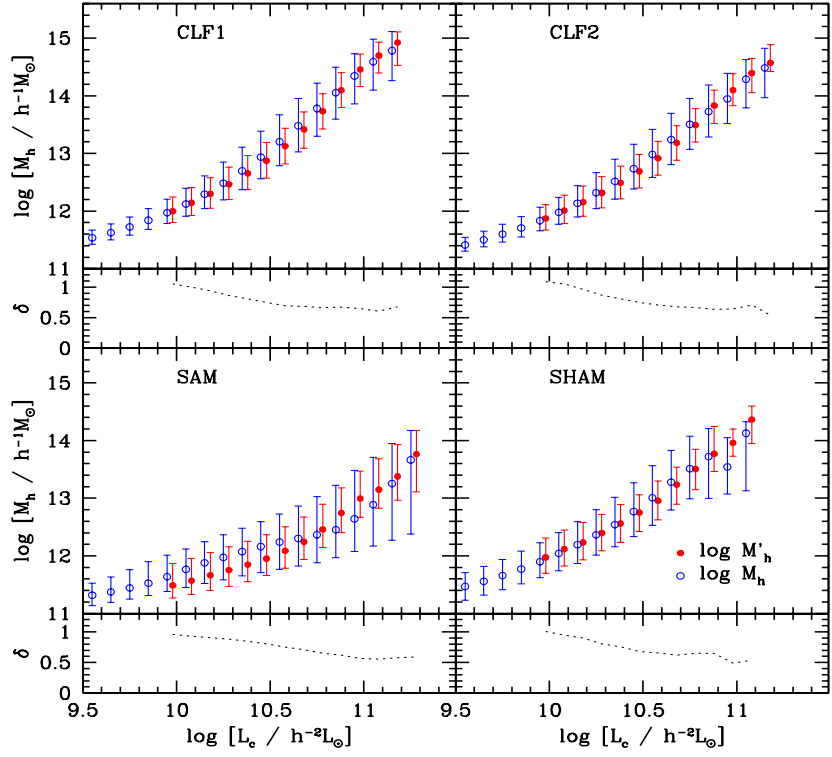}
\caption{Similar to the results shown in Fig.  \ref{fig:Lc-Mh}, but
  here we compare the original (open circles) $\log M_h$ and
  luminosity gap pre-corrected (solid circles) $\log M'_h$ halo
  masses.  Obviously, the latter has significantly reduced scatters in
  $\log M_h$, especially at the massive end. The ratio between these
  two error bars, $\delta$, are shown in the small panels with dashed lines. }
\label{fig:M_correct}
\end{figure*}

\begin{figure*}
\center
\vspace{0.5cm}
\includegraphics[height=10.0cm,width=11.0cm,angle=0]{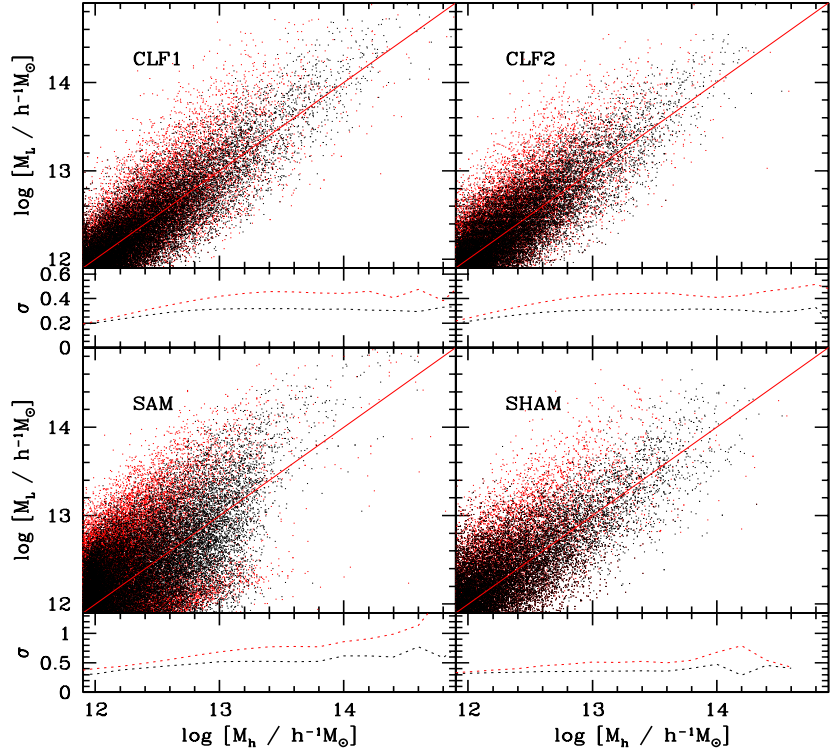}
\caption{Comparison to the true halo mass for four mock samples. Red
  points are halo masses obtained using Eq. \ref{eq:Mh-Lc}, while
  black points are obtained using Eq. \ref{eq:M_func}.  Red solid
  lines show the relation when $\log M_L$ equals to $\log M_h$.
  The standard variance of galaxies from the red solid lines,
  $\sigma$, are shown in the small panels. The red dotted line
  corresponds to red points of the above panels, while black one
  corresponds to black points. }
\label{fig:CompareMass}
\end{figure*}

To model halo mass using also luminosity  gap, we formally write
\begin{equation}
\label{eq:M_func}
\log M_h(L_c,L_{\rm gap}) = \log M_h(L_c) + \Delta \log M_h(L_c,L_{\rm gap}) \,,
\end{equation}
where the first term on the right side is the empirical relation
described by Eq. (\ref{eq:Mh-Lc}), while $\Delta \log M_h(L_c,L_{\rm
  gap})$ represents the amount of correction to the $M_h(L_c)$
relation with the help of $L_{\rm gap}$.  Fig.  \ref{fig:DeltaM-gap}
shows the relation between luminosity gap and $\Delta \log M_h$ for
given central galaxy luminosity, where $\Delta \log M_h$ is defined by
the ratio between the true halo mass and that predicted by
eq. (\ref{eq:Mh-Lc}). Shown in different panels are for halos
(groups) in different $L_c$ bins, centered at $\log L_c = 9.9$, 10.2,
10.5, 10.8, with a bin width equal to 0.1, respectively.

The red solid points represent the median $\Delta \log M_h$ for each
luminosity gap subsamples constructed from the CLF1 sample, with error
bars indicating the 68 percentile. In all cases $\Delta \log M_h$
decreases with increasing luminosity gap, consistent with the trend
seen in Fig \ref{fig:Mh-gap}, where one sees that halos with a smaller
luminosity gap are more massive.  The overall amplitude of $\Delta
\log M_h$ increases with the central luminosity, consist with the
growing sizes of errorbars at the large $L_c$ end shown in Fig.
\ref{fig:Lc-Mh}.  For different $L_c$ the asymptotic values of $\Delta
\log M_h$ are also different.

We use the following functional form to model $\Delta \log
M_h(L_c,L_{\rm gap})$,
\begin{equation}\label{eq:DM_func}
\Delta \log M_h(L_c,L_{\rm gap}) = \eta_a \exp(\eta_b \log L_{\rm gap}) + \eta_c\,,
\end{equation}
where parameters $\eta_a$, $\eta_b$ and $\eta_c$ may all depend on
$L_c$.  The parameter $\eta_b$ in general has a negative value which
describes the decline rate of $\Delta \log M_h$ with $\log L_{\rm
  gap}$.  Parameter $\eta_c$ describes the asymptotic value at large
$\log L_{\rm gap}$, while parameter $\eta_a+\eta_c$ is the value of
$\Delta \log M_h$ at $\log L_{\rm gap}=0$. To gain some idea of the
functional forms of these three parameters, we first fit to the data
for each of the $\log L_c$ bins shown in Fig.\,\ref{fig:DeltaM-gap},
and obtain the values of $\eta_a$, $\eta_b$ and $\eta_c$.  According
to their general dependence on $\log L_c$, we use the following forms to
model these three parameters as functions of $\log L_c$:
%eq5
\begin{eqnarray}\label{eq:eta_abc}
\eta_a(L_c)~ &=&~ \exp(\log L_c-\beta_1) \nonumber \\
\eta_b(L_c)~ &=&~ \alpha_2(\log L_c -\beta_2) \,, \\
\eta_c(L_c)~ &=&~ -(\log L_c - \beta_3)^{\gamma_3} \nonumber
\end{eqnarray}
which in total has five free parameters.  We fit all the data for CLF1
(as shown partly in Fig. \ref{fig:DeltaM-gap}) in the luminosity range
$10.9 \ge \log L_c \ge 9.7$ with the above model using a Monte-Carlo
Markov Chain (MCMC) to explore the likelihood function in the
  multi-dimensional parameter space (see Yan, Madgwick \& White 2003;
  van den Bosch et al.  2005b for more detail).  Then, we chose the
  parameter set has the highest likelihood (minimum $\chi^2$) to be
  the model parameters. The best fit values together with their 68\%
  confidence ranges are listed in the first row of Table
  \ref{tab:DeltaM}. This set of parameter values are valid for
central galaxies in the luminosity range $10.9 \ge \log L_c \ge 9.7$.
For central galaxies outside this range, one can apply the correction
factor at $\log L_c = 9.7$ for fainter galaxies and $\log L_c = 10.9$
for brighter galaxies.  At the faint end, the asymptotic correction
factor $\Delta \log M_h$ at $\log L_c = 9.7$ is small.  At the bright
end, the reason to cut at $\log L_c = 10.9$ is that our data for
massive halos are statistically poor and not included in the fitting
and that the fitting function for the $M_h(L_c)$ relation (Eq.
\ref{eq:Mh-Lc}) starts to deviate from data at $\log L_c > 10.9$.

\begin{deluxetable*}{lccccc}
 \tabletypesize{\scriptsize} \tablecaption{Parameters for the $\Delta \log
   M_h$ function} \tablewidth{0pt}

  \tablehead{$\Delta \log M_h$ & $\beta_1$  & $\alpha_2$   & $\beta_2$   &
    $\beta_3$   & $\gamma_3$ } \\
  \startdata

  MEMBER 2 & $    10.37^{+ 0.23}_{- 0.25}$ &  $     2.14^{+ 4.22}_{- 1.81}$ &  $   -11.57^{
+ 0.64}_{- 6.40}$ &  $     9.90^{+ 0.02}_{- 0.16}$ &  $     3.29^{+ 4.13}_{- 1.5
4}$    \\
   \\
  MEMBER 3 &  $    10.09^{+ 0.16}_{- 0.19}$ &  $     0.25^{+ 1.50}_{- 0.15}$ &  $   -16.51^{
+ 5.29.46}_{- 6.62}$ &  $     9.86^{+ 0.04}_{- 0.20}$ &  $     2.92^{+ 2.00}_{- 1.1
2}$    \\
   \\
  MEMBER 4 & $    9.84^{+ 0.15}_{- 0.22}$ &  $     0.15^{+ 0.71}_{- 0.07}$ &  $   -16.70^{
+ 5.18}_{- 3.90}$ &  $     9.69^{+ 0.10}_{- 0.21}$ &  $     2.82^{+ 0.87}_{- 0.6
3}$     \\
    \\
  MEMBER 5 &  $    9.71^{+ 0.13}_{- 0.18}$ &  $     0.10^{+ 0.43}_{- 0.02}$ &  $   -17.74^{
+ 6.07}_{- 0.16}$ &  $     9.60^{+ 0.10}_{- 0.19}$ &  $     2.75^{+ 0.76}_{- 0.3
5}$      \\
    \\
  MEMBER 6 & $    9.57^{+ 0.11}_{- 0.24}$ &  $     0.07^{+ 0.29}_{- 0.01}$ &  $   -18.02^{
+ 6.14}_{- 0.39}$ &  $     9.49^{+ 0.10}_{- 0.24}$ &  $     2.77^{+ 0.40}_{- 0.2
6}$     \\
   \\
  MEMBER 7 & $     9.78^{+ 0.13}_{- 0.30}$ &  $     -0.89^{+ 0.75}_{- 0.70}$ &  $   -9.59^{
+ 2.39}_{- 0.25}$ &  $     9.80^{+ 0.10}_{- 0.40}$ &  $     2.64^{+ 1.59}_{- 0.7
4}$     \\
    \\
  MEMBER 8 & $     9.53^{+ 0.33}_{- 0.12}$ &  $     -0.28^{+ 0.13}_{- 1.10}$ &  $   -8.84^{
+ 1.00}_{- 1.00}$ &  $     9.47^{+ 0.43}_{- 0.16}$ &  $     2.55^{+ 1.02}_{- 0.5
0}$    \\
    \\
  MEMBER 9 & $     9.91^{+ 0.11}_{- 0.20}$ &  $     0.07^{+ 0.05}_{- 0.01}$ &  $   -16.78^{
+ 2.46}_{- 0.46}$ &  $     9.38^{+ 0.07}_{- 0.14}$ &  $     2.63^{+ 0.19}_{- 0.1
8}$    \\

 \enddata
\label{tab:DeltaM}
\end{deluxetable*}

\begin{figure*}
\center
\vspace{0.5cm}
\includegraphics[height=10.0cm,width=11.0cm,angle=0]{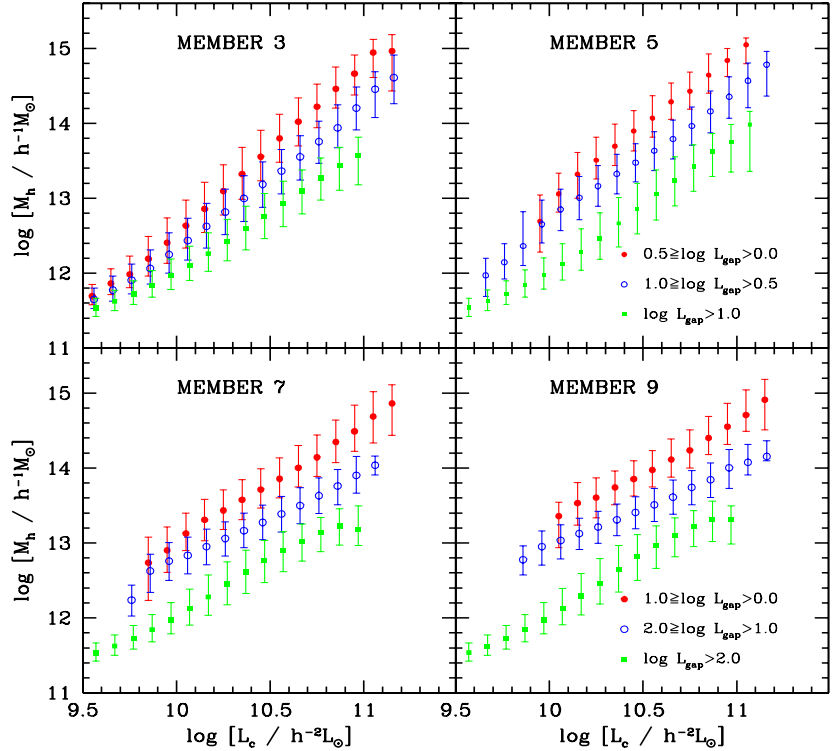}
\caption{Similar to the results shown in Fig.  \ref{fig:Mh-gap}, but
  here for CLF1 mock sample with three, five, seven and nine brightest
  member galaxies been observed and used in calculating the luminosity
  gap.  Groups within three different luminosity gap ranges are shown
  using different symbols as indicated in the upper and lower right
  panels.  Here again, symbols and error bars represent the median and
  68\% confidence levels of $\log M_h$ in different $\log L_c$ bins. }
\label{fig:Mh-gap-order}
\end{figure*}

The best fit $\Delta \log M_h$ as a function of $\log L_{\rm gap}$ is
shown in each panel of Fig. \ref{fig:DeltaM-gap} as the solid line.
Obviously, the model describes the overall behavior in $\Delta \log
M_h(L_c,L_{\rm gap})$ remarkably good.  For comparison, we also show
$\Delta \log M_h(L_c,L_{\rm gap})$ as a function of $\log L_{\rm gap}$
obtained from the CLF2, SAM and SHAM samples in each panel of
Fig.\ref{fig:DeltaM-gap} using different symbols as indicated.
Although coming from completely different galaxy formation models
and/or cosmological parameters, the halo mass correction factors for
other samples are quite similar to that for the CLF1 sample.  Such
agreement indicates that our model for $\Delta\log M_h(L_c,L_{\rm
  gap})$ is quite insensitive to the details of galaxy formation and
cosmology.

Since the main purpose of this paper is to obtain a better halo mass
estimator, we compare the amount of scatter in the original $\log
M_h(L_c)$ relation and the new $\log M_h(L_c,L_{\rm gap})$ model.  To
perform such a comparison, we define a `pre-corrected' halo mass
\begin{equation}\label{eq:M'_h}
\log M'_h = \log M_h - \Delta \log M_h(L_c,L_{\rm gap})\,,
\end{equation}
and check if the scatter in the $M'_h(L_c)$ relation is significantly
reduced relative to that in the $M_h(L_c)$.  If the correction were
perfect, the scatter in the $M'_h(L_c)$ would be reduced to 0.

Fig. \ref{fig:M_correct} demonstrates the performance of the corrected
model.  Here results for the original $M_h(L_c)$ relations are shown
as the open circles and the $M'_h(L_c) $ relation as the solid points.
The error-bars are all 68 percentiles around the median values in the
$L_c$ bins.  It is clear that the scatter in $M'_h(L_c)$ is
significantly reduced relative to the original relation, especially
for massive halos/groups where the scatter is reduced by about 50\%.
For clarity, we define the ratio between the corrected and
  original error bars as $\delta$ in the sub-panels. We can see,
  $\delta$ close to 1.0 at low mass ends indicate that two error bars
  is about the same.  Then $\delta$ drop to about 0.5 for four samples
  show that the corrected error bar can reduce 50\% of the original
  one.  Note that the correction model $\Delta \log M_h$ is
calibrated by the CLF1 sample alone, but its application to other
samples also leads to significant reduction in the scatter.  These
results indicate clearly that adding of even just the second brightest
galaxy can give a significant improvement of the mass estimation over
using the central (the brightest) galaxy alone.

In order to further test the reliability of this halo mass estimation,
we compared the halo mass $\log M_L$ obtained from the galaxy
  luminosity using Eq. \ref{eq:M_func} to the true halo mass $\log
  M_h$ for each group which are directly available in our four mock
  samples: CLF1 CLF2, SAM and SHAM galaxy catalogs.
Fig. \ref{fig:CompareMass} shows this comparison. The red points
represent the halo masses estimated using only $L_c$, i.e., using
Eq. \ref{eq:Mh-Lc}. While the black points represent the halo masses
estimated using both $L_c$ and $L_{\rm gap}$ (Eq. \ref{eq:M_func}). In
order to quantify the scatter with respect to the line of equality
($\log M_L = \log M_h$) in each panel, we
measure the standard variance $\sigma$ between estimated and true halo
mass for each galaxy group.  The $\sigma$ is defined as:
\begin{equation}\label{eq: sigma}
\sigma = \sqrt{\frac{\sum_{i=1}^n(\log M_h-\log M_L)^2}{n-1}}\,,
\end{equation}
where $n$ is the number of groups in each halo mass bin with width
$\Delta \log M_h=0.1$.  The results, shown in the small panels,
with red dotted lines represent the standard variance for only using
$L_c$ and black ones represent the estimation by using both $L_c$ and
$L_{\rm gap}$, indicate that the estimations for halo mass are all
improved for four mock samples.  Overall, adding the second brightest
member galaxy can roughly reduce the scatter by $\sim 0.3$dex at the
massive end.

We also noticed that the halo masses estimated using $L_c$ alone has
somewhat systematical deviation from the true halo masses at the
massive end. Here the systematic deviation is mainly induced by the so
called Eddington bias where there are significantly more low mass
halos than massive halos that can scattered to larger ones.  As we can
see, with the help of the second brightest member galaxy, this
deviation decreases substantially. And we will see later that using
additional member galaxies, such systematic deviation indeed
disappears finally.

\subsection{Using subsequent member galaxies}
\label{richness}

Having demonstrated that the luminosity gap between central and the
brightest satellite galaxy can be used to better constrain the halo
mass estimation, we come to subsequent member galaxies. Since the
richness is also an indicator for halo mass and correlated with $L_c$
(Shen et al. 2014), we may expect the luminosity gap between central
and the other satellite galaxies also contains information of halo
mass.  In this subsection, we will extend our study to fainter member
satellite galaxies. Like before, we still use the `luminosity gap'
$L_{gap} = L_c/L_s$, but here $L_s$ is the luminosity of $i$-th member
galaxy in consideration with $L_s = L_3$ or $L_4$, etc., where $L_3$
and $L_4$ are the luminosities of the third and fourth brightest
galaxies in this group.  Note that, in general one can also define
$L_s$ as the total luminosity of all member satellite galaxies in
consideration. However, as we have tested, this definition dose not
improve any of our results. While the gap thus defined has negative
values. For simplicity, we adopt $L_s$ as luminosity of the $i$-th
brightest member galaxy.

\begin{figure*}
\center
\vspace{0.5cm}
\includegraphics[height=17.0cm,width=17.0cm,angle=0]{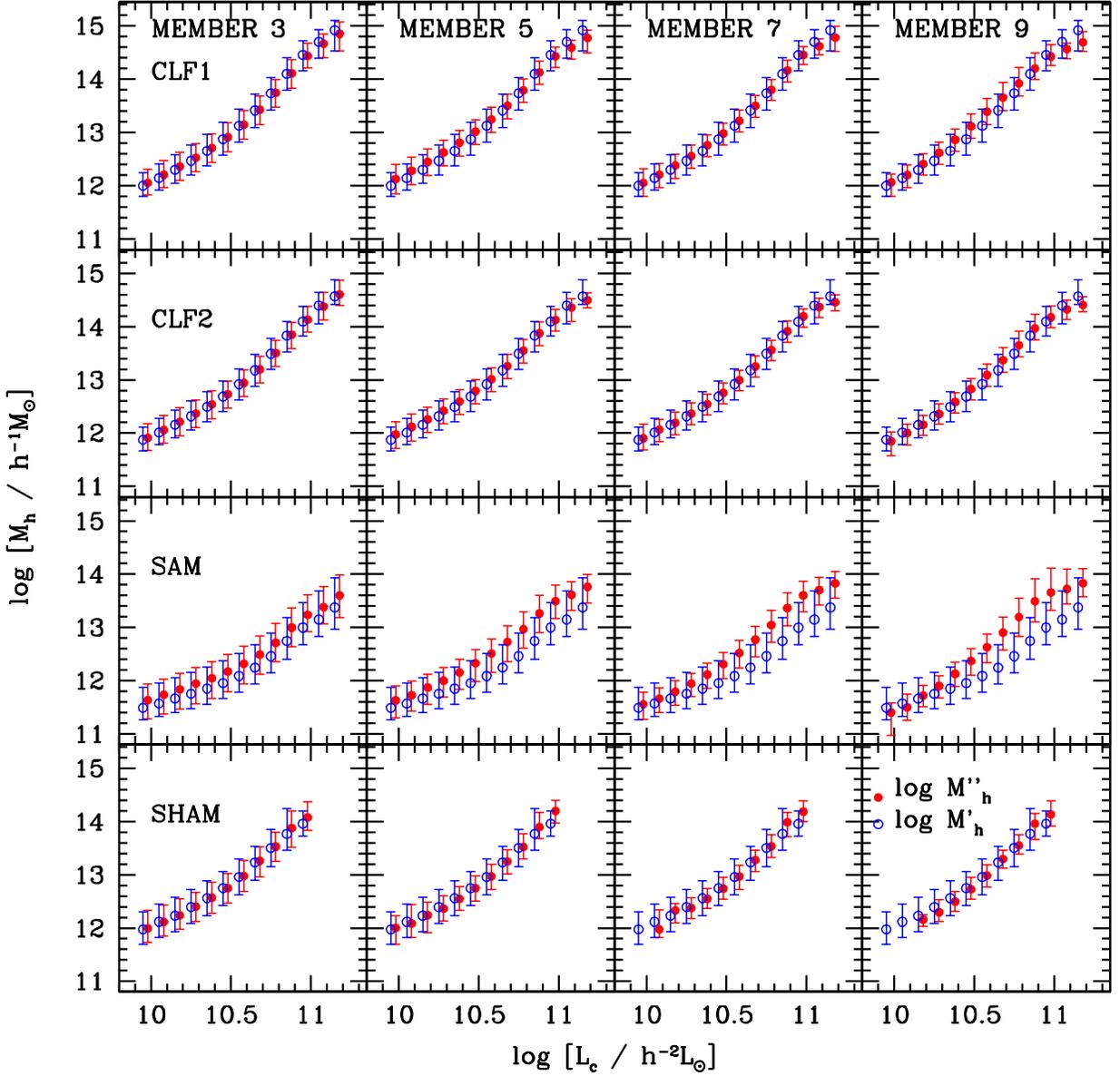}
\caption{Similar to the results shown in Fig.  \ref{fig:M_correct} but
  here for four mock samples (CLF1, CLF2, SAM and SHAM from top to
  bottom rows) with three, five, seven and nine brightest members
  galaxies been observed and used in calculating the luminosity gap.
  Here we compare the luminosity gap pre-corrected (open circles) halo
  mass $\log M'_h$ (the second brightest galaxy) in Fig.  \ref{fig:M_correct} with the
  ones obtained from the third, fifth, seventh and ninth brightest member
  galaxies, respectively.  Obviously, the latter make further
  reduction on scatters in $\log M''_h$, especially at the massive
  end.  }
\label{fig:M_correct-CLF}
\end{figure*}

We start with the third brightest member galaxy and stop at the
ninth. For the fainter member galaxy, the luminosity gap will be
larger.  Overall, the values of $L_{gap}$ are roughly in the range
$0.0 - 3.0$.  As before, we split groups into three categories
according to the value of the luminosity gap. The separation criteria
for groups with member less than five are: small gap ($0.5 \ge \log
L_{\rm gap}> 0.0$), median gap ($1.0 \ge \log L_{\rm gap} > 0.5$) and
large gap ($\log L_{\rm gap} > 1.0$). The criteria for groups with six
and more member galaxies are: small gap ($1.0 \ge \log L_{\rm
  gap}>0.0$), median gap ($2.0 \ge \log L_{\rm gap} > 1.0$) and large
gap ($\log L_{\rm gap} > 2.0$).  As an illustration, we show in Fig.
\ref{fig:Mh-gap-order} the gap-dependent halo mass distributions, here
for CLF1 sample only.  From top left to right bottom are the results
of luminosity gap between central and the third, fifth, seventh and
ninth brightest members, respectively.  Within each panel only halos
which contain at least the indicated number of member galaxies with
$\log L>8.0$ are used.  In all four panels, the relationships between
central galaxy luminosity and the halo mass are similar, and the
overall luminosity gap dependence is similar to that shown in Fig.
\ref{fig:Mh-gap} for the $M_h(L_c)$ relations.  Compare to Fig.
\ref{fig:Mh-gap}, the separation of halo mass between different
luminosity gaps increases along with the richness. For groups using
more than five member galaxies, the variance between different gaps
are more distinct than that by using three or four member galaxies,
which suggest that a better constraint on the halo mass can be
obtained by using fainter member galaxies, as we will see below.

\begin{figure*}
\center
\vspace{0.5cm}
\includegraphics[height=10.0cm,width=11.0cm,angle=0]{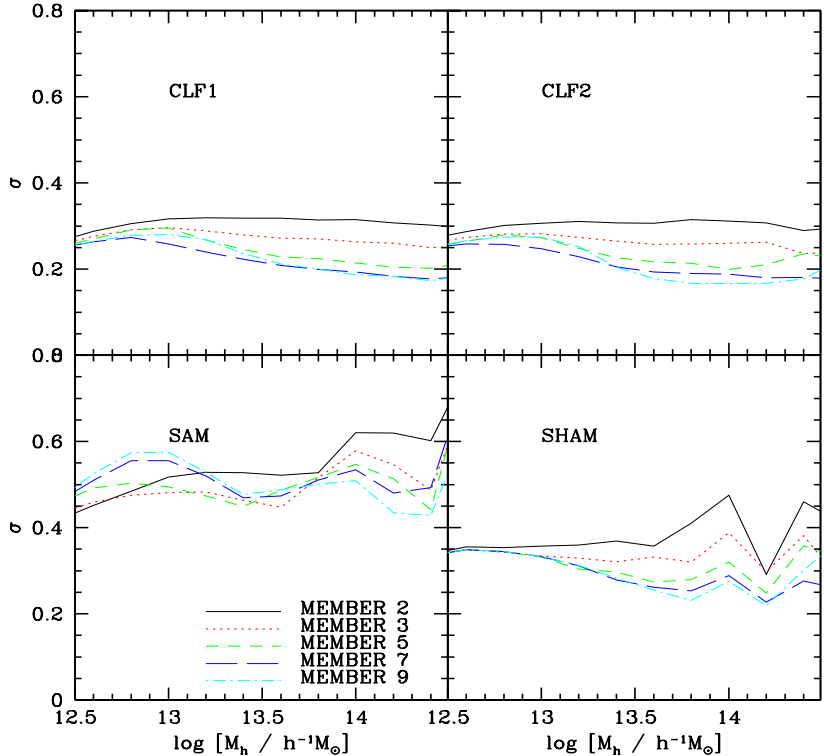}
\caption{Similar to the results shown in Fig. \ref{fig:CompareMass},
  but here we only show the $\sigma$ values for four samples: CLF1,
  CLF2, SAM and SHAM. Lines with different colors represent different
  member galaxies that are applied to calculate the luminosity
  gap. }
\label{fig:CompareMassObs}
\end{figure*}

For a given number of members starting from three, we first repeat
calculations similar to those shown in Fig. \ref{fig:DeltaM-gap}, and
then split the sample into several bins according to the central
galaxy luminosity, starting from $\log L_c = 9.8$ to $11.0$ with bin
width equal to 0.1.  The $\Delta \log M_h(L_c, L_{\rm gap}) - \log
L_{\rm gap}$ relations in the luminosity range $11.0 \ge \log L_c \ge
9.8$ are used to obtain the best fit parameters that describe the
corrections defined in Eq.\,(\ref{eq:M_func}), using the same MCMC
algorithm as to the central-second cases. The parameters so obtained
are listed in the second to eighth row of Table \ref{tab:DeltaM} for
cases using the third, fourth, $\cdot\cdot\cdot$, ninth brightest
members, respectively. This set of parameters are again only valid for
central galaxies within the luminosity range used in fitting.  For
central galaxies with luminosities outside this region, one can apply
the correction factors at $\log L_c = 9.8$ for faint galaxies and at
$\log L_c = 11.0$ for bright galaxies.  We find that the functional
form (Eq. 12) describes the overall $\Delta \log M_h(L_c, L_{\rm
  gap})$ remarkably well for every central galaxy luminosity bins
considered, and for all cases regardless which member galaxy used.

Note that the parameters listed in Table \ref{tab:DeltaM} are obtained
from the CLF1 sample alone.  To see if the same correction model can
also be applied to other samples, we have checked the $\Delta \log
M_h$ distributions for the CLF2, SAM and SHAM samples as well, and
found that the model also works well.

To show the performance of the correction model, we use the same
definition of `pre-corrected' halo mass $\log M'_h$ as described in
Eq. (\ref{eq:M'_h}). To distinguish with the one obtained from the
second brightest member galaxy, we denote it with $\log M''_h$. As
demonstrated in Fig. 5, the scatter is significantly reduced in the
$M'_h(L_c)$ relation when the second brightest galaxies are used in
the halo mass estimate.  By using subsequent fainter members, we
expect the model to be improved further.  In Fig.
\ref{fig:M_correct-CLF} we compare the `pre-corrected' halo mass as a
function of $L_c$ between cases using different member galaxies
obtained from the CLF1, CLF2, SAM and SHAM samples in different row
panels as indicated.  The blue open circles with error bars show the
median and 68 percentile of $\log M'_h$, while the solid squares with
error bars show the same quantities of $\log M''_h$.

We see that, in all four samples, involving fainter member galaxies do
make further reduction in the scatter of $M_h(L_c)$ relation compared
to using the second brightest galaxies.  The correction method works
more effectively for groups with larger central galaxy luminosity
($\log L_c \ge 10.5$). Overall there is about 40\% additional
reduction in the scatter at the massive end for a fainter (e.g. the
fifth - seventh brightest) member galaxy.  For lower luminosities
($10.5 \ge \log L_c \ge 10.0$), the scatter itself shows that
including the second or more fainter member galaxies ($\log M'_h$ v.s.
$\log M''_h$) does not change the scatter significantly, because the
halo mass is dominated by the central galaxy itself, and the scatter
is already quite small even without the correction.

To better assess the quality of our halo mass estimation with
additional member galaxies, we check the difference between the
extracted and true halo masses of the groups similar to that shown in
the small panels of Fig. \ref{fig:CompareMass}.  We show in
Fig. \ref{fig:CompareMassObs} the standard variance, $\sigma$
(Eq. \ref{eq: sigma}), of estimated halo mass from true halo mass of
each group. In each panel lines with different colors represent the
estimations by using the second, third, fourth $\cdot\cdot\cdot$,
ninth brightest members of CLF1, CLF2, SAM and SHAM mock samples,
respectively. In general, as expected, using fainter member galaxies
does improve the performance of the estimation, especially for massive
halos. Interestingly, we also noticed that this improvement is not a
monotonically growth with the number of member galaxies we
concerned. When the number of member galaxies involved is larger than
five, the amounts of improvement actually are quite limited and become
negligible when using more than seven members. This feature is
  caused by the decomposition of our halo mass estimation method: the
  median $L_c-M_h$ relation together with a gap dependent correction.
  This decomposition has the advantage of shrinking the halo mass
  estimation error only in very poor groups. As we will show in the
  next section, for groups with more than five members, `RANK' method
  still performs better given good survey condition. Hence, we treat
the seventh member galaxies as the most appropriate option in this
paper. In future applications, we will use halo mass estimation by
involving seven member galaxies at most.

\begin{figure*}
\center
\vspace{0.5cm}
\includegraphics[height=17.0cm,width=17.0cm,angle=0]{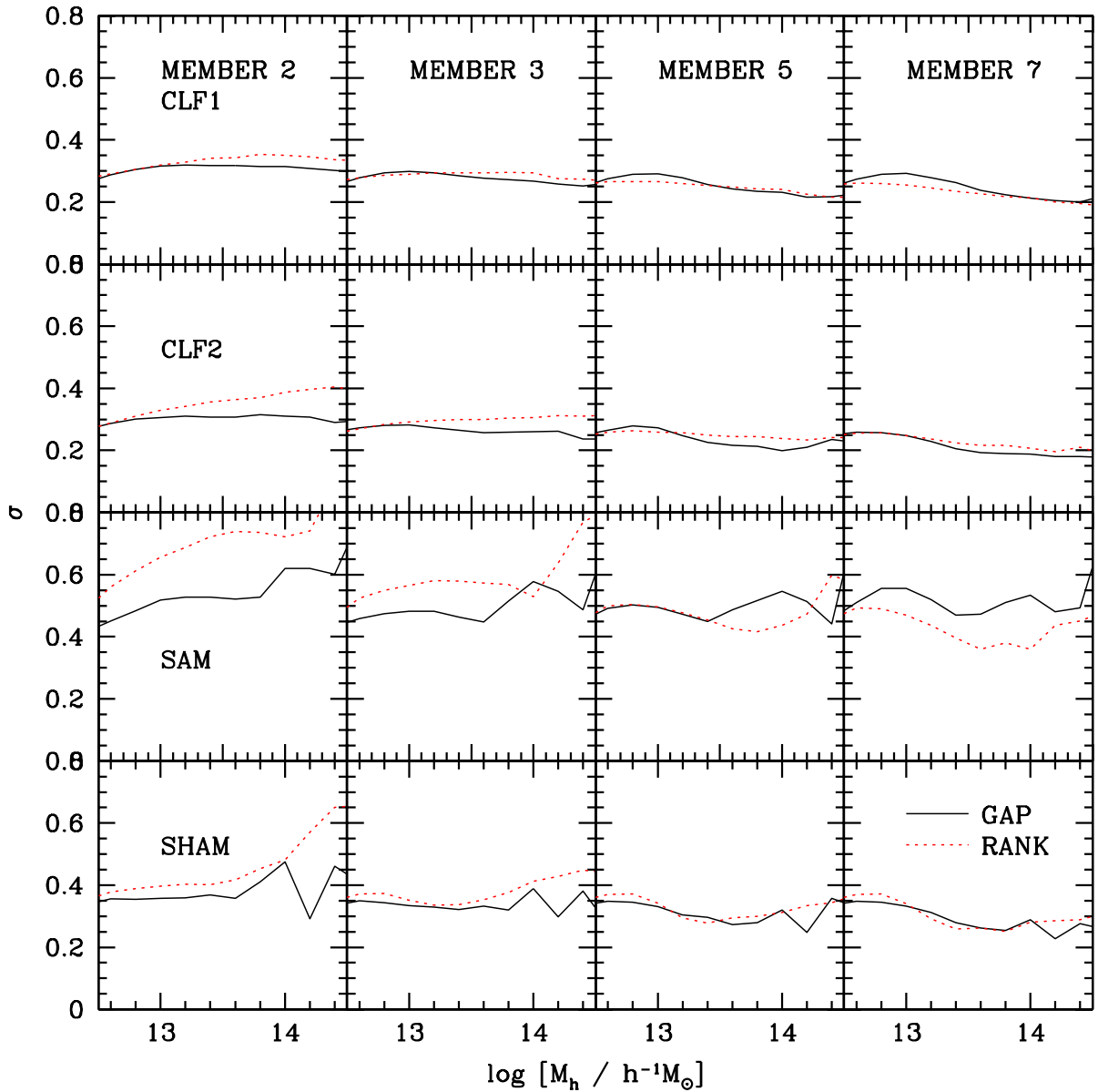}
\caption{The comparison of $\sigma$ for two halo mass estimation
  methods which involve two, three, five and seven brightest member
  galaxies in each galaxy group of CLF1, CLF2, SAM and SHAM mock
  samples, respectively.  The black solid lines represent the `GAP'
  method, which are the same as those shown in Fig.
  \ref{fig:CompareMassObs}.  Red dotted lines represent the results for the
  `RANK' method which are described in the \S \ref{sec
    richness}. }
\label{fig:sigmaMn}
\end{figure*}

\begin{figure*}
\center
\vspace{0.5cm}
\includegraphics[height=10.0cm,width=11.0cm,angle=0]{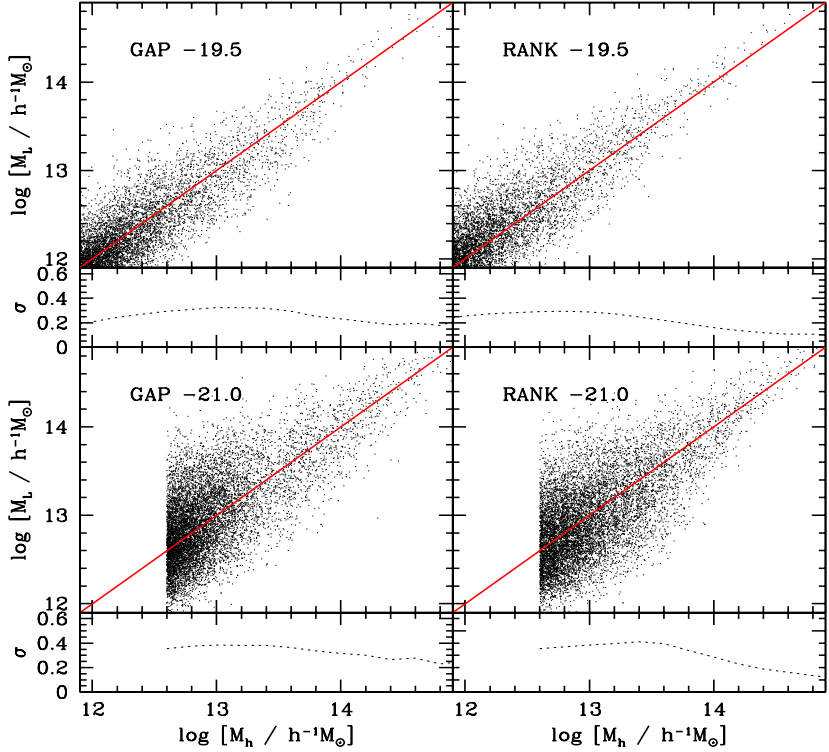}
\caption{Similar to Fig. \ref{fig:CompareMass}, but here the
  comparisons are carried out between `GAP' and `RANK' methods based
  on CLF1 mock sample with two different absolute magnitude cuts:
  $\rmag=-19.5$ (top) and -21.0 (bottom). }
\label{fig: Mmagcut}
\end{figure*}

\begin{figure*}
\center
\vspace{0.5cm}
\includegraphics[height=7.0cm,width=11.0cm,angle=0]{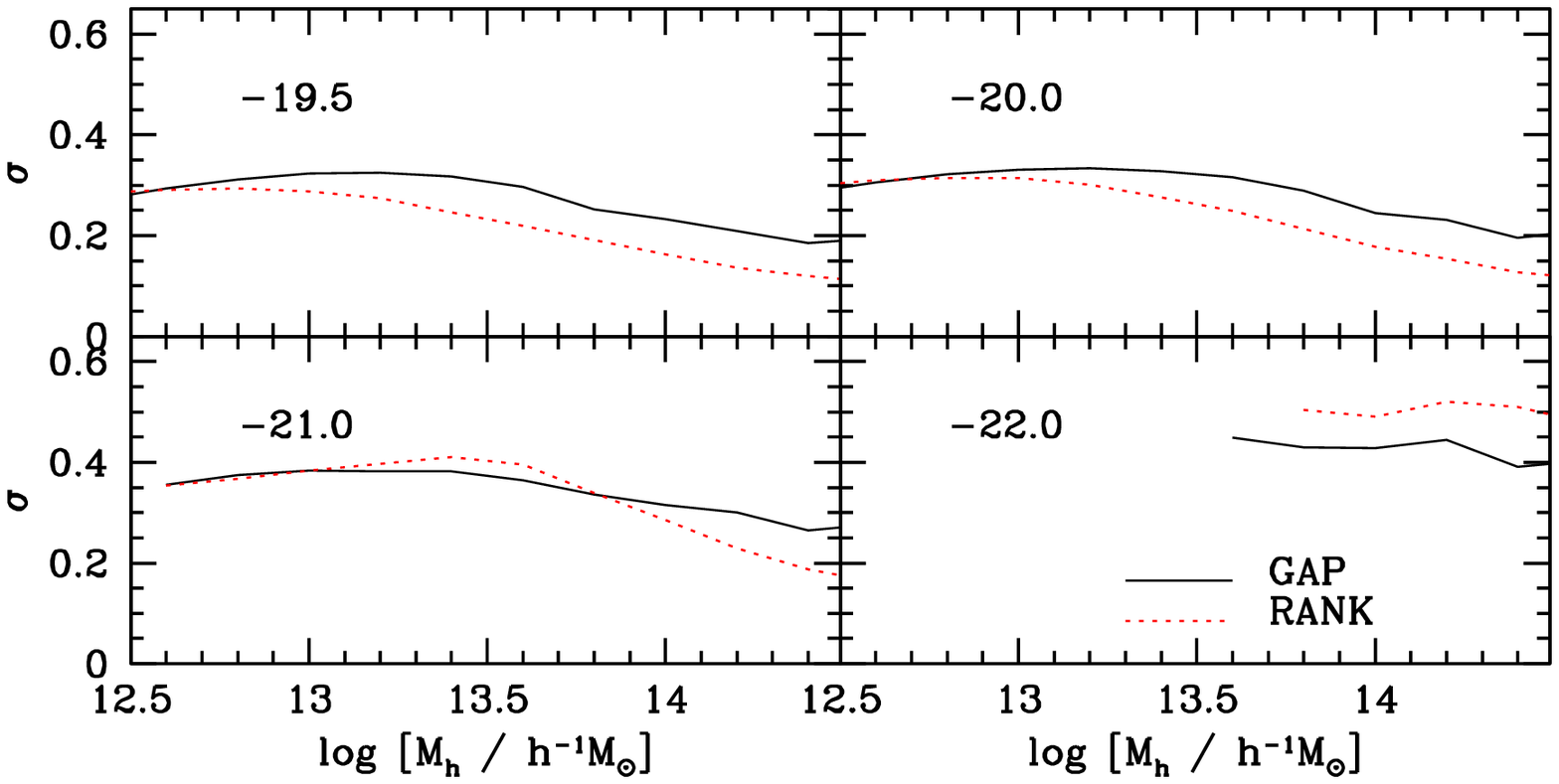}
\caption{Similar to Fig. \ref{fig: Mmagcut}, but here only show the
  $\sigma$ values for both `GAP' (red line) and `RANK' (black line)
  methods for four absolute magnitude cuts: $\rmag=-19.5$, -20.0,
  -21.0 and -22.0, respectively. }
\label{fig: Msigma}
\end{figure*}

\section{Comparing with the ranking method}
\label{sec rank}

Having demonstrated the ability of improving the halo mass estimations
using luminosity gap between the central and satellite galaxies, we
would like to see if this correction model already achieves the
similar good performance of estimating halo mass using the ranking of
$L_G$ (see Fig. \ref{fig:M-Lg}). Hereafter we refer these two methods
as `GAP' and `RANK', respectively.

\subsection{The effect of richness}
\label{sec richness}

The results from last section show the performance of `GAP' method
somehow depends on how many member galaxies are involved. Meanwhile,
this difference of richness certainly will impact on the performance
of `RANK' method which is based on the characteristic group luminosity
of member galaxies. Therefore, it would be interesting to compare the
performances of `GAP' and `RANK' methods on group systems of same
richnesses. To make fair comparisons between groups of the same
richness, we update the characteristic group luminosity $L_G$ with
$L'_G$, which is defined as the total luminosity of the brightest
member galaxies to be included. For example, when we only involve
three member galaxies, then $L'_G = L_1 + L_2 + L_3$. Then, the
estimation of halo mass is based on the ranking of $L'_G$ instead of
$L_G$.

Fig.\ref{fig:sigmaMn} shows the results of such a comparison. Similar
to Fig. \ref{fig:CompareMass} and Fig. \ref{fig:CompareMassObs},
$\sigma$ is the standard variance of estimated halo masses from true
halo masses. In each panel, the solid black line is given by the `GAP'
method and is the same as that shown in Fig. \ref{fig:CompareMassObs},
which uses the second, third, fifth and seventh brightest members for
CLF1, CLF2, SAM and SHAM mock samples, respectively. The dotted red
line in each panel is given by the `RANK' method using only two, three,
five and seven brightest member galaxies.

In general, the performance of the `GAP' method is somewhat better
when only a few (less than five) brightest member galaxies are
considered. Then, this advantage is gradually lost when involving more
members. In the row only five member galaxies are considered, the
performance of `GAP' and corrected `RANK' method is about the
same. When the number of member galaxies reach to seven, the `RANK'
method tends to be slightly better especially for the SAM sample.
Meanwhile, we could also see the impact of including different
richness for each galaxy group on `RANK' method is larger than `GAP'
method. Thus, `GAP' method would be a better option when only a few
galaxy members can be obtained, especially for groups with members
less than five.

\subsection{The effect of magnitude limit}

In real observations, one can only observe galaxies brighter than the
survey magnitude limit. This limiting magnitude in-turn allows us to
have a complete observation of galaxies of given absolute magnitude to
certain redshift. Here we make use of CLF1 mock sample to assess the
performances of the `RANK' and `GAP' methods under different absolute
magnitude cuts. Note that, for `GAP' method, there are several ways to
calculate luminosity gap depending on which member galaxy are used. As
we found in previous section that the `GAP' method roughly reaches the
best performance at 7th brightest member galaxy, therefore, in what
follows, when we refer to `GAP' method, we mean using the maximum
available up to 7th member galaxies to estimate the halo masses.

Fig. \ref{fig: Mmagcut} shows the comparisons between true halo masses
directly obtained from the mock samples and the ones estimated using `GAP'
(left panels) and `RANK' (right panels) methods, respectively. We
chose the absolute magnitude cuts to be $\rmag=-19.5$ (top) and
$\rmag=-21.0$ (bottom). Using galaxies brighter than these absolute
magnitudes in a group, we estimate the halo masses of the groups
according to their characteristic group luminosity and luminosity gap
respectively.  In general, the performances of `RANK' is slightly
better than `GAP' method, especially with a fainter absolute magnitude
cut. Once we go to brighter absolute magnitude cut, both the `RANK'
and `GAP' methods have worse accuracy of halo mass estimation.  The
amount of variance between the true halo masses and the estimated
ones increase by about 0.1 dex from $\rmag=-19.5$ to $\rmag=-21.0$ for
both estimation methods.  Nevertheless, we see that the performance of
`GAP' method is approaching to that of the `RANK' method.  Here,
compare to results show in Fig.\ref{fig:CompareMass}, we do see
that the systematic deviation in mass estimation disappears.

To reveal in more detail the impact of different magnitude cuts, we
plot the comparison of $\sigma$ for both `RANK' and `GAP' estimation
methods in Fig.\ref{fig: Msigma} to much brighter absolute magnitude
cuts: $\rmag=-19.5$, -20.0, -21.0 and -22.0.  These absolute
  magnitude cuts roughly correspond to redshift completeness limit
  $z=0.09$, 0.103, 0.157, 0.22 in the SDSS observation with apparent
  magnitude limit $r=17.6$ (e.g. Wang et al.  2007). The black lines
in Fig.\ref{fig: Msigma} represent the results given by `GAP' method,
while red lines are given by 'RANK' method.  Apparently, missing faint
member galaxies influences both methods.  Compare to `GAP' method,
`RANK' method is more sensitive to the increasing of luminosity
threshold. In $\rmag=-19.5$ panel, `RANK' method has an obvious
advantage over a wide halo mass range, while the situation is inverted
at $\rmag=-22.0$.  These results may indicate that, for very high
redshift or very shallow observations where the luminosity threshold
is high, `GAP' method may be a better option. In addition, the `GAP'
method is not suffered from the volume calculation which is required
in the `RANK' method, thus can be easily implemented to surveys
  with poor geometry and are flux limited.

\section[]{SUMMARY}
\label{sec_conclusion}

Galaxies are thought to form and reside within code dark matter
  halos.  Different galaxy formation processes, e.g., star forming,
  quenching, AGN feedback, etc., may occur or dominate in halos of
  different masses.  On the other hand, the halos are the building
  block of the cosmic web, one can use the halo mass functions and
  clustering properties to understand the nature of our
  Universe. Within these studies, halo mass estimation from various
  observations is one of the key challenges. However, such a mission
is difficult, especially for poor galaxy systems where only a few
brightest member galaxies are observed.  Within the galaxy formation
framework, the central (brightest) galaxies in the dark matter halos
have a rough scaling relation with the masses of the halos: brighter
galaxies live in more massive halos.  However, the scatter within that
scaling relation is very large, especially in massive halos. In this
study, we make use of four mock galaxy catalogues, one based on the
SAM, one based on the SHAM and the other two on the CLF, to
investigate the central-halo scaling relation and to find out a way
that can significantly reduce its scatter.  Based on these four mock
samples, we probed the impact of the luminosity gap, which is defined
as the luminosity ratio between the brightest and other member
galaxies in the same dark matter halo.  In this paper, we take into
account a maximum of nine member galaxies in the modeling. We find
that the scatter in the central-host halo mass relation is luminosity
gap dependent, which in turn can be used to reduce this intrinsic
scatter. The main findings of this paper are as follows.
\begin{enumerate}
\item We find that the median halo mass for a given central galaxy
  luminosity can be described by simple relation described by
  Eq. \ref{eq:Mh-Lc}, however with quite large scatter around this
  median.
\item The scatter in the halo mass depends both on the central galaxy
  luminosity and the luminosity gap between the central and the
  subsequent brightest member galaxies.
\item We have obtained a mass correction factor Eq. \ref{eq:DM_func}
  which is independent to the detailed galaxy formation models, and
  thus can be applied to any median halo mass - central galaxy
  luminosity ($M_h-L_c$) relation to get better estimation of the halo
  masses.
\item The correction factors can reduce the scatters in halo mass
  estimations in massive halos by about 50\% to 70\% depend on which
  member (second or seventh) galaxies are used.
\item Comparing this `GAP' method with traditional `RANK' method, we
  find that the former performs better for groups with less than five
  members, or in observations with very bright magnitude cut.  In
  addition, the `GAP' method does not need to calculate the volume in
  estimating halo masses, and thus is much easier to be applied to
  observations with very small volume or with poor geometry.
\end{enumerate}

The above modeling is very useful for our probing of those poor galaxy
systems or shallow observations where only a few brightest member
galaxies in a halo can be observed.  Based on these limited member
galaxies, we can have a fairly good estimation of the halo mass, which
is important for various astrophysical and cosmological studies.

\section*{Acknowledgements}

We thank the anonymous referee for helpful comments that greatly
improved the presentation of this paper. We also thank H.J. Mo and
Frank C. van den Bosch for useful discussions.  This work is supported
by the 973 Program (No. 2015CB857002), NSFC (Nos. 11128306, 11121062,
11233005), the Strategic Priority Research Program ``The Emergence of
Cosmological Structures" of the Chinese Academy of Sciences, Grant No.
XDB09000000, and a key laboratory grant from the Office of Science and
Technology, Shanghai Municipal Government (No. 11DZ2260700).

%%%%%%%%%%%%%%%%%
% Appendix
%%%%%%%%%%%%%%%%%

\appendix

\section{A different definition of luminosity gap}

\begin{figure*}
\center
\vspace{0.5cm}
\includegraphics[height=10.0cm,width=11.0cm,angle=0]{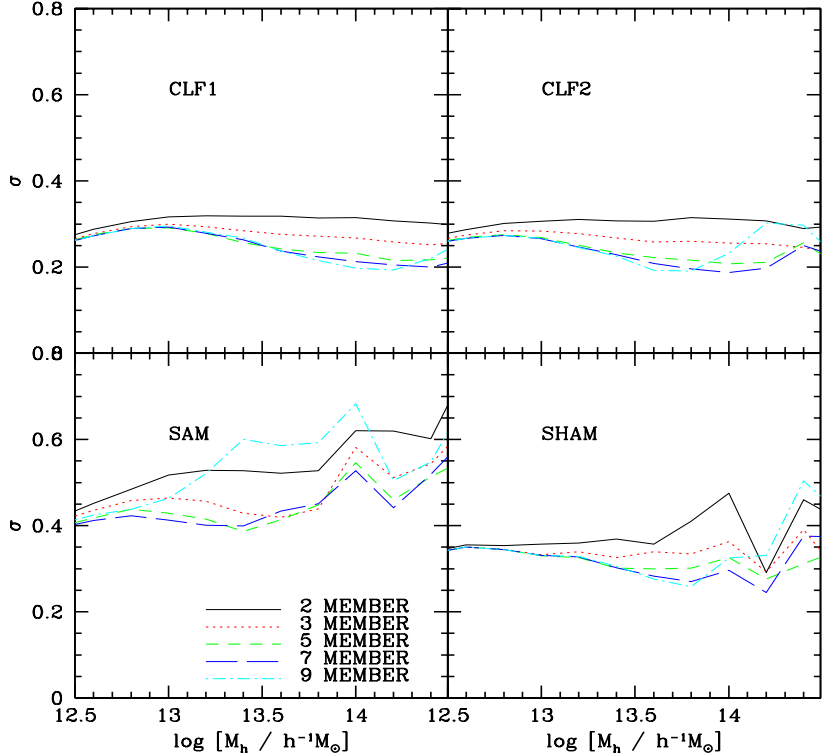}
\caption{Similar to the results shown in Fig. \ref{fig:CompareMassObs},
  but here the $\sigma$ values are given by another definition of
  luminosity gap for CLF1,CLF2, SAM and SHAM sample respectively.
  Lines with different colors represent different number of
  member galaxies that are applied to calculate the luminosity
  gap. }
\label{fig:AF1}
\end{figure*}

Throughout the paper, we define the luminosity gap to be the
  difference between the brightest and the $i$-th brightest member
  galaxies. In general, we can also define the luminosity gap as the
  difference between the brightest and {\it all} satellite galaxies,
  $\log L_{\rm gap} = \log L_c - \log L_s$, here $\log L_s$ is defined
  as the total satellite galaxy luminosity in consideration. For
  example, if we only consider two brightest member galaxies in a
  group, $\log L_s = L_2$. If we choose to include four member
  galaxies in a group, then $\log L_s = L_2+L_3+L_4$, where $L_3$ and
  $L_4$ are the luminosities of the third and fourth brightest
  galaxies in this group.

Then, we applied exactly the same procedure as before using this
  new luminosity gap definition. First, calculate the $\Delta \log
  M_h(L_c, L_{\rm gap}) - \log L_{\rm gap}$ relations as in
  Fig. \ref{fig:DeltaM-gap} in the luminosity range $10.9 \ge \log L_c
  \ge 9.7$. Then, use MCMC algorithm to find the best parameter set
  which can describe the relation defined by eq.  \ref{eq:DM_func} and
  eq. \ref{eq:eta_abc}. Note that, since we applied total luminosity
  of member galaxies, the luminosity gap $\log L_{\rm gap}$ can be
  negative. Overall, the values of $\log L_{\rm gap}$ are roughly in
  the range $-1.0 - 2.0$. Using the relation obtained by the new
  definition of luminosity gap, we estimated the halo mass for each
  group in four mock samples and compared these estimated halo mass to
  the true halo mass using $\sigma$ as before.  Fig. \ref{fig:AF1}
  shows the performance of this new definition of luminosity
  gap. Compare to those shown in Fig. \ref{fig:CompareMassObs}, we can
  see the tendency and value of $\sigma$ are quite similar in the two
  figures.  In general, the definition for $i$-th member galaxy
  luminosity is slightly better at massive ends.

\label{lastpage}

\end{document}